\documentclass[twocolumn,prb,showpacs,preprintnumbers,superscriptaddress,amsmath,amssymb,citeautoscript]{revtex4-1}
\usepackage{graphicx}
\usepackage{epstopdf}
\usepackage{dcolumn}
\usepackage{color}
\usepackage{amsmath,amssymb}
\usepackage[english]{babel}
\usepackage{bm}
\usepackage[hyperindex]{hyperref}
\hypersetup{colorlinks,citecolor=blue, filecolor=blue,linkcolor=blue,urlcolor=blue}

\graphicspath{{./figures/}}
\usepackage{multirow}
\usepackage{rotating}
\usepackage{array}

\newcommand{\bpm}{\begin{pmatrix}}
\newcommand{\epm}{\end{pmatrix}}
\newcommand{\bs}{\boldsymbol}
\newcommand{\be}{\begin{equation}}
\newcommand{\ee}{\end{equation}}
\newcommand{\beq}{\begin{eqnarray}}
\newcommand{\eeq}{\end{eqnarray}}

\DeclareMathOperator{\sgn}{sgn}
\DeclareMathOperator{\re}{Re}
\DeclareMathOperator{\im}{Im}

\DeclareMathOperator{\tr}{tr}

\begin{document}

\title{Surface Green's functions and boundary modes using impurities: Weyl semimetals and topological insulators}
\author{Sarah Pinon}
\affiliation{Institut de Physique Th\'eorique, Universit\'e Paris Saclay, CEA
CNRS, Orme des Merisiers, 91190 Gif-sur-Yvette Cedex, France}
\author{Vardan Kaladzhyan}
\email{vardan.kaladzhyan@phystech.edu}
\affiliation{Department of Physics, KTH Royal Institute of Technology, Stockholm, SE-106 91 Sweden}
\author{Cristina Bena}
\affiliation{Institut de Physique Th\'eorique, Universit\'e Paris Saclay, CEA
CNRS, Orme des Merisiers, 91190 Gif-sur-Yvette Cedex, France}

\date{\today}

\begin{abstract}
In this work we provide a new direct and non-numerical technique to obtain the surface Green's functions for three-dimensional systems. This technique is based on the ideas presented in Phys. Rev. B \textbf{100}, 081106(R), in which we start with an infinite system and model the boundary using a plane-like infinite-amplitude potential. Such a configuration can be solved exactly using the T-matrix formalism. We apply our method to calculate the surface Green's function and the corresponding Fermi-arc states for Weyl semimetals. We also apply the technique to systems of lower dimensions, such as Kane-Mele and Chern insulator models, to provide a more efficient and non-numerical method to describe the formation of edge states.
\end{abstract}

\maketitle

\section{Introduction}

Boundaries of certain condensed matter systems host unique phenomena. For instance, graphene exhibits zero-energy zigzag-edge modes \cite{Fujita1996}, and topological insulators exhibit conducting edge or surface states \cite{Kane2005, Bernevig2013, Hsieh2008}. In order to describe boundary effects, several techniques were developed, including the exact diagonalization of tight-binding Hamiltonians \cite{Slater1954, Busch1987}, iterative methods to compute boundary Green's functions \cite{Sancho1984, Sancho1985, Peng2017}, solving the Schr\"{o}dinger equation \cite{Duncan2018} and the bulk-boundary correspondence \cite{Rhim2018}. 

A new method describing the formation of boundary modes was recently introduced\cite{Kaladzhyan2018}. This method can be generalized to any dimensions, and in certain situations it can yield fully analytical results, providing a deeper physical insight than numerical techniques. The general idea is as follows: instead of considering a finite system with a sharp boundary, we consider an infinite system with a strong delta-potential impurity emulating the shape of the boundary. For example, in order to recover end, edge or surface boundaries, the impurity potential should be chosen to be point-like, line-like and plane-like, respectively. In the limit of an infinite impurity potential such impurities divide a given system into two independent semi-infinite regions. Subsequently we use the $T$-matrix formalism \cite{Bena2016, Balatsky2006} to study the impurity-induced states which transform into boundary states when the impurity strength is larger than any energy scale in the system. 

\begin{figure}
\centering
\includegraphics[width=0.9\columnwidth]{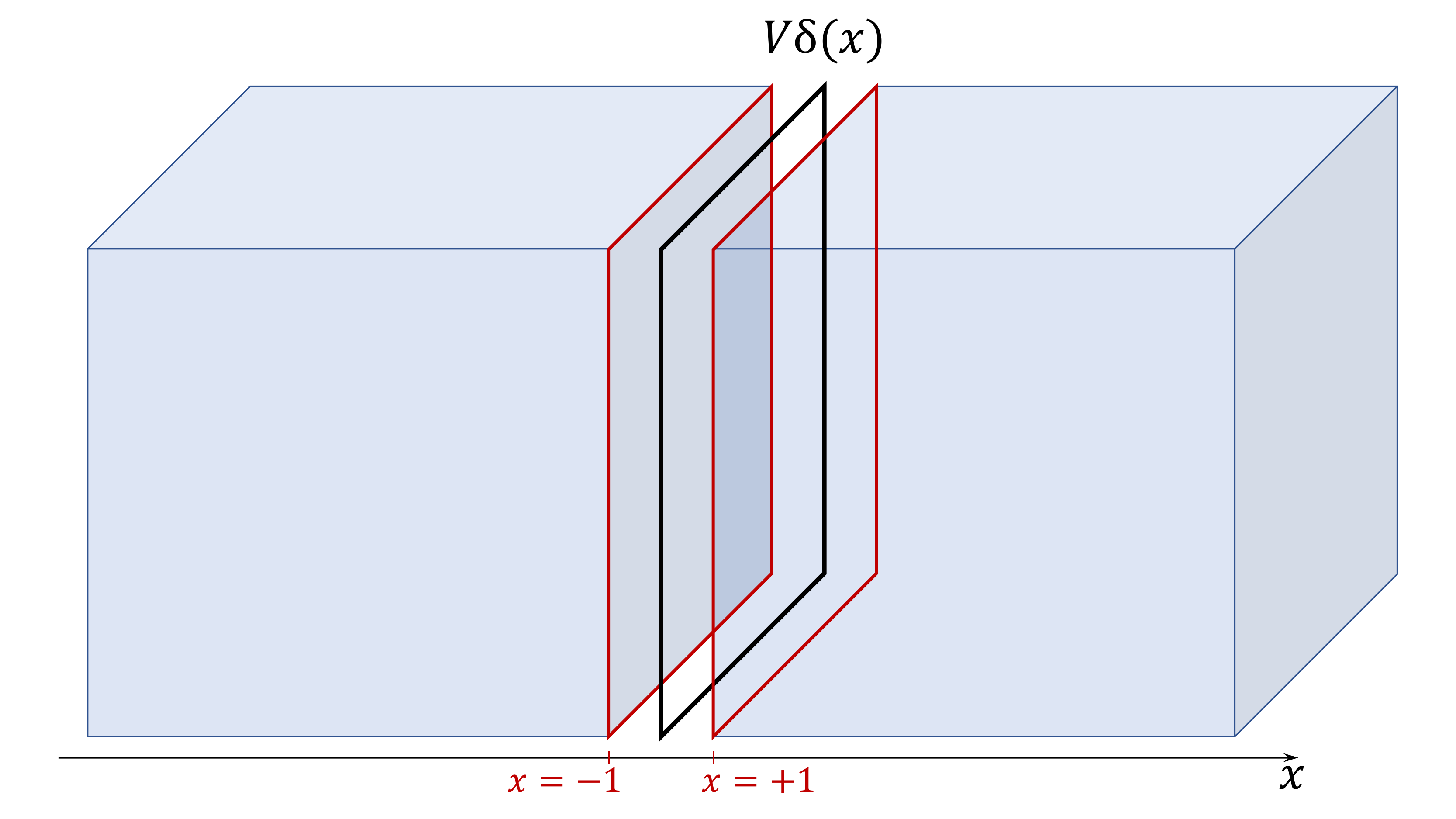}
\caption{Schematics of the 3D systems and surface GFs. The black parallelogram is the impurity plane, while the red ones show the two created surfaces on the neighboring planes at $x = \pm 1$, one lattice constant away from the impurity plane.}
\label{fig:sys}
\end{figure}

Along the same lines, we present here a direct and non-numerical technique to calculate the surface Green's functions of an arbitrary three-dimensional system. The boundary can once more be modeled as a plane impurity potential with an amplitude going to infinity. The corresponding full Green's functions can be calculated exactly using the $T$-matrix formalism. The resulting Green's function evaluated on the plane neighboring and parallel to the impurity plane becomes the surface Green's function (see Fig.~\ref{fig:sys}). We apply this technique to calculate the surface Green's functions for Weyl semimetals described by two different models\cite{Kourtis2016,Lau2017}. We recover in each case the corresponding Fermi-arc states.

Moreover, in this work we apply the technique from Ref.~[\onlinecite{Kaladzhyan2018}] to a new class of systems -- topological insulators. In particular we consider a 2D honeycomb lattice described by the Kane--Mele model\cite{Kane2005}, as well as a 2D Chern insulator \cite{Bernevig2013}. We show that impurity-induced states in these two models transform into helical or chiral edge modes, respectively, when the impurity potential is taken to infinity. While the Kane--Mele model requires performing a numerical integration, the 2D Chern insulator allows an exact closed-form solution and thus demonstrates the analytical power of the method.

The paper is organized as follows: in Sec.~II we introduce the formalism and the notations. In Sec.~III we present the calculation of the surface Green's functions for a Weyl semimetal described by two different models, and the formation of the corresponding Fermi-arc states. In section IV and V we focus on two-dimensional topological insulators described by the Kane--Mele and Chern-insulator models, respectively, and we obtain the corresponding edge modes. We leave the conclusions to Sec.~VI. 

\section{$T$-matrix formalism for surface Green's functions and edge states}

Below we consider an infinite system described by a momentum-space Hamiltonian $\mathcal{H}_{\bs{k}}$. The unperturbed Matsubara Green's function (GF) can be written as: $G_0\left(\bs{k},i\omega_n\right) = \left[i\omega_n - \mathcal{H}_{\bs{k}}\right]^{-1}$, where $\omega_n$ denote the Matsubara frequencies. In the presence of an impurity, the Green's function is modified to:
\begin{align}
\label{eq:green_perturbed}
G\left( \bs{k}_1 ,\bs{k}_2,i\omega_n \right) &= G_0\left(\bs{k}_1,i\omega_n\right) \delta_{\bs{k}_1,\bs{k}_2} \\ 
&+ G_0\left(\bs{k}_1,i\omega_n\right) T\left(\bs{k}_1,\bs{k}_2,i\omega_n\right) G_0\left(\bs{k}_2,i\omega_n\right) \nonumber 
\end{align}
where the $T$-matrix $T\left(\bs{k}_1,\bs{k}_2,i\omega_n\right)$ embodies all-order impurity-scattering processes \cite{Balatsky2006, Mahan2000}. Note that due to the impurity-induced breaking of translational symmetry, and consequently of the momentum conservation, the generalized Green's function depends no longer on one, but on two values of momentum. For a delta-function impurity $V_{\mathrm{imp}}\left(\bs{r}\right) \equiv V \delta \left(x\right)$, the form of the $T$-matrix in 1D is momentum independent and is given by \cite{Balatsky2006, Ziegler1996, Salkola1996, Bena2008}:
\begin{align}
\label{eq:T_matrix1D}
T\left(i\omega_n\right) = \left[\mathbb{I} - V \cdot \int \frac{dk_x}{L_k} G_0\left(k_x,i\omega_n\right)\right]^{-1} \cdot V
\end{align}
while in 2D and 3D, for a line and plane impurity respectively, localized at $x=0$ and perpendicular to the $x$ axis, we have:
\begin{align}
\label{Tmatrix2D} 
T&\left(k_{1y},k_{2y},i\omega_n\right)= \\
&=\delta_{k_{1y},k_{2y}} \left[\mathbb{I} - V \cdot \int \frac{d k_x}{L_k} G_0\left(k_x,k_{1y},i\omega_n\right)\right]^{-1} \cdot V \nonumber
\end{align}
and
\begin{align}
\label{Tmatrix3D} 
\nonumber T&\left(k_{1y},k_{1z},k_{2y},k_{2z},i\omega_n\right)= \delta_{k_{1y},k_{2y}} \delta_{k_{1z},k_{2z}} \times \\
&\times\left[\mathbb{I} - V \cdot \int \frac{dk_x}{L_k} G_0\left(k_x,k_{1y},k_{1z},i\omega_n\right)\right]^{-1} \cdot V,
\end{align}
respectively, with $L_k$ being a normalization factor. The limits of integration are given by the boundaries of the first Brillouin zone, i.e., for a 1D system and for a 2D square lattice or 3D cubic lattice we integrate from $-\pi$ to $\pi$ (with $L_k = 2\pi$), while for a honeycomb lattice with an impurity along $y$ we integrate  from $-2\pi/3$ to $2\pi/3$ ($L_k = 4\pi/3$) (for a full justification see Appendix \ref{FT}). Note that Eqs.~(\ref{Tmatrix2D}) and~(\ref{Tmatrix3D}) are independent of $k_{1x}$ and $k_{2x}$ due to the fact that the impurity potential is a delta-function centered at $x=0$. Reversely, note that the T-matrix contains the terms $\delta_{k_{1y},k_{2y}}$  (2D) and $\delta_{k_{1y},k_{2y}}\delta_{k_{1z},k_{2z}}$ (3D), since the impurity is independent of $y$ in 2D and of $y$ and $z$ in 3D, and therefore, the momenta in the corresponding directions are conserved in all scattering processes.

The exact same formalism can be applied for impurities perpendicular to the other axes of the systems.

In what follows we employ this formalism at zero temperature to calculate the retarded GF $\mathcal{G}(\bs{k}_1, \bs{k}_2, E)$ obtained by the analytical continuation of the Matsubara GF $G(\bs{k}_1, \bs{k}_2, i\omega_n)$ (i.e., by setting $i\omega_n \rightarrow E + i\delta$, with $\delta \rightarrow 0^+$). 

For a three-dimensional system the surface Green's function can be extracted from the perturbed generalized Green's function in Eq.~(\ref{eq:green_perturbed}). This can be related to the mixed Green's function in which we keep the momentum coordinates in the two directions parallel to the impurity plane ($k_y$ and $k_z$), but we perform a Fourier transform to write down the Green's function in real space coordinates in the $x$ direction.  Note that for a plane impurity the generalized Green's function depends on two different values of momentum only for the direction perpendicular to the impurity, in the other two directions we recover a simple dependence on momentum due to unbroken translational invariance. Thus we have:
\begin{align}
&\mathcal{G}_s(k_{y},k_{z}) \equiv \mathcal{G}(x=x'=\pm1;k_{y},k_{y};k_{z},k_{z})&\nonumber\\
&=\int \negthickspace \frac{dk_{1x}}{L_k}\int\negthickspace \frac{dk_{2x}}{L_k}\mathcal{G}(k_{1x},k_{2x};k_{y},k_{y};k_{z},k_{z}) e^{i k_{1x} x}e^{-i k_{2x} x'}.&
\end{align}
We fix $x=x'=\pm1$ since we are interested in describing the lattice planes one lattice constant away from the impurity (see Fig.~\ref{fig:sys}).  The boundaries of the two resulting semi-infinite systems correspond to the two planes at $x=\pm1$, and the Green's functions taken at these two positions are effectively the surface Green's functions for the semi-infinite systems. The physics at $x=0$ (impurity position) is relatively trivial since the infinite-amplitude impurity potential pushes away all the wave function weight off the impurity plane. For simplicity we have omitted writing down explicitly the energy dependence of the Green's functions. Note again that translational invariance holds within the planes parallel to the impurity plane, thus the surface Green's functions depend only on one momentum in each of the in-plane directions. Furthermore, in order to obtain the surface physics the value of the impurity potential in Eq.~(\ref{Tmatrix3D}) needs to be set to a value much larger than all the energy scales in the problem.

The surface Green's function allows us to recover the formation of the surface states such as, for instance, the Fermi-arc states. Thus, we can study the surface spectral function 
\begin{equation}
\label{eq:spectral_function_s}
A(k_y,k_z,E) = -\frac{1}{\pi} \im\{\tr[\mathcal{G}_s\left(k_y,k_z,E\right) ]\}.
\end{equation} 

The same analysis can be performed for a two-dimensional system with a line-impurity to find the line Green's functions 
\begin{align}
&\mathcal{G}_l(k_{y})=\mathcal{G}(x=x'=\pm1;k_{y},k_{y})&\nonumber\\
&=\int \frac{dk_{1x}}{L_k}\int \frac{dk_{2x}}{L_k}\mathcal{G}(k_{1x},k_{2x};k_{y},k_{y}) e^{i k_{1x}x}e^{-i k_{2x}x'}.&
\end{align}
and the corresponding edge states given by:
\begin{equation}
\label{eq:spectral_function_l}
A(k_y,E) = -\frac{1}{\pi} \im\{\tr[\mathcal{G}_l\left(k_y,E\right) ]\}.
\end{equation} 

Alternatively, in order to visualize the impurity-induced states, as described in Appendix \ref{kxintegral}, we may focus on the average correction to the spectral function:
\begin{equation}
\delta N(k_y,E) =  \int \frac{dk_x}{L_k} \delta A(k_x,k_y,E).
\label{def}
\end{equation}
where
\begin{equation}
\label{eq:spectral_function}
\delta A(\bs{k},E) = -\frac{1}{\pi} \im\{\tr[\mathcal{G}_0\left(\bs{k}\right) T\left(\bs{k},\bs{k}\right) \mathcal{G}_0\left(\bs{k}\right)]\}.
\end{equation} 
Above $\mathcal{G}_0\left(\bs{k}\right)$ stands for $\mathcal{G}_0\left(\bs{k},E\right)$ and $T\left(\bs{k},\bs{k}\right)$  for $T\left(\bs{k},\bs{k},E\right)$.
The integral over $k_x$ is performed along the same interval as the one defined in Eq.~(\ref{Tmatrix2D}). This quantity corresponds to the average number of available electronic states with wavevector $(k_y,k_z)$, where the average is performed along the direction perpendicular to the impurity. A more detailed description of the significance of this quantity is provided in Appendix \ref{kxintegral}.

\section{Weyl semimetals}
In what follows we consider a Weyl semimetal: there is a large number of models of various degree of complexity describing such a system. Here we focus only on the tight-binding models described in Refs.~[\onlinecite{Kourtis2016}] and [\onlinecite{Lau2017}], which we denote by $\mathcal{H}_1$ and $\mathcal{H}_2$, respectively. The Bloch Hamiltonians for these two systems are given by
\begin{eqnarray}
H_{1,2}=\sum_{\bs{k}} \psi^\dagger(\bs{k}) \mathcal{H}_{1,2}(\bs{k})\psi(\bs{k}),
\end{eqnarray}
where $\psi(\bs{k})=(c_{\bs{k} A \uparrow},c_{\bs{k} A \downarrow}, c_{\bs{k} B \uparrow}, c_{\bs{k} B \downarrow})$ is a spinor with the index $A/B$ denoting a generic unspecified orbital component, and the $ \uparrow/ \downarrow$ the physical spin.

For the model in Ref.~[\onlinecite{Kourtis2016}] written in the basis above we have
\begin{eqnarray}
\mathcal{H}_{1}(\bs{k})&=&g_1(\bs{k}) \tau_1\sigma_3+g_2(\bs{k}) \tau_2 \sigma_0+g_3(\bs{k})\tau_3\sigma_0
\nonumber\\&&+g_0(\bs{k})\tau_0\sigma_0+\beta \tau_2\sigma_2+\alpha \sin k_y \tau_1\sigma_2,
\end{eqnarray}
where 
\begin{eqnarray}
g_0(\bs{k})&=&2d(2-\cos k_x-\cos k_y)\nonumber \\
g_1(\bs{k})&=&a \sin k_x\nonumber \\
g_2(\bs{k})&=&a \sin k_y\nonumber \\
g_3(\bs{k})&=&m + t \cos k_z+ 2b(2-\cos k_x-\cos k_y).
\end{eqnarray}
and $\alpha$, $\beta$ are real parameters. The $2\times 2$ identity matrices $\sigma_0/\tau_0$ and the Pauli matrices $\sigma_i/\tau_i$, $i=1,2,3$ act in the spin and the orbital spaces, correspondingly, and the multiplication of the $\sigma$ and $\tau$ matrices indicates a tensor product.

We consider the same values of parameters as those in Ref.~[\onlinecite{Kourtis2016}], thus we take a) $a = b = 1$, $t = -1$, $m = 0.5$, $d = 0.8$, $\alpha=\beta=0$ and b) $a = b = 1$, $t = -1.5$, $d = m = 0$, $\beta = 0.9$, and $\alpha = 0.3$. The former is characterized by two Weyl points, while the latter by four Weyl points, and thus we expect to have one and two Fermi arcs, respectively. 

\begin{figure}
\centering
\vspace*{-1cm}
\hspace*{-0.5cm}
\includegraphics[width=9.6cm]{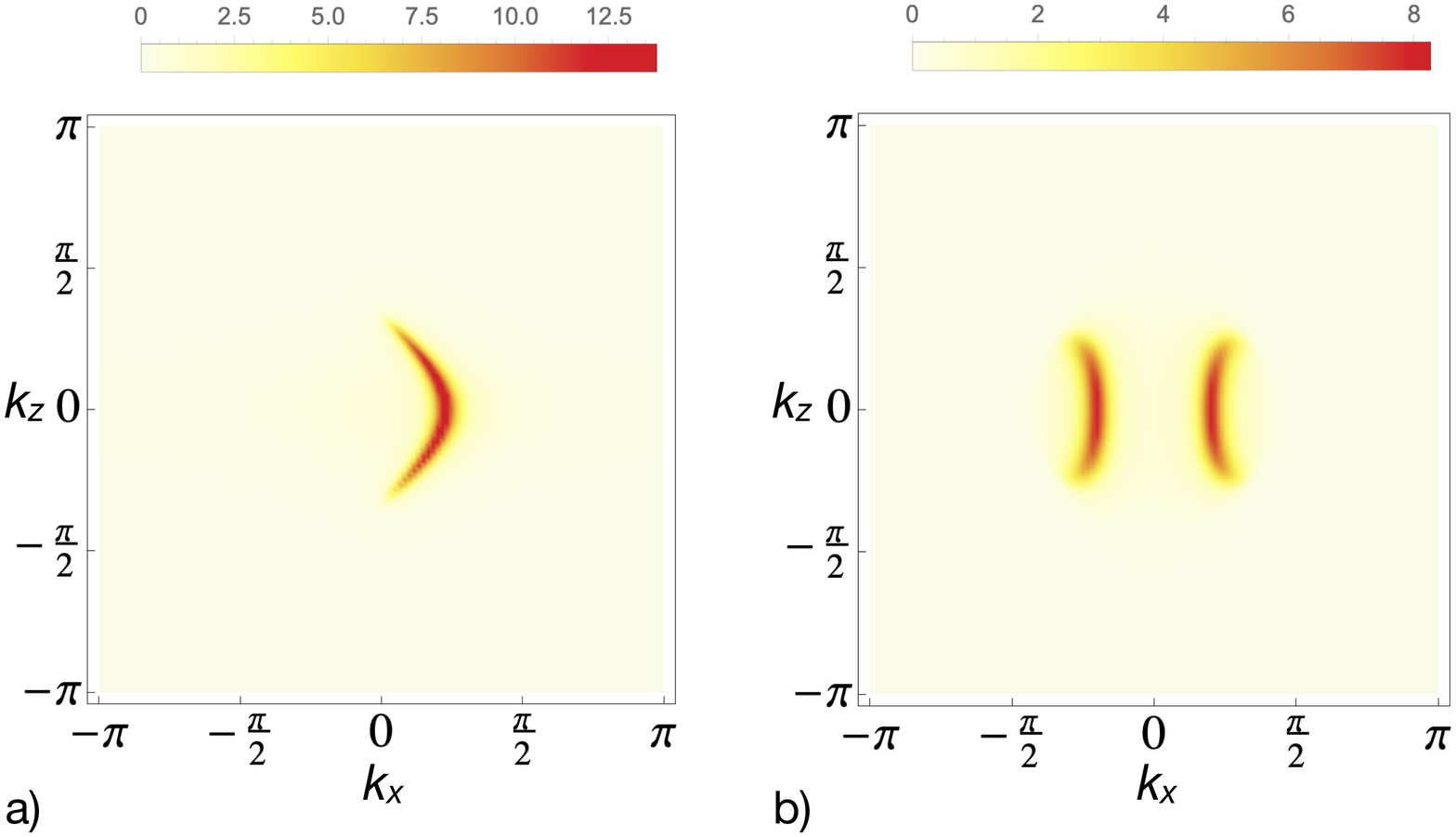}
\vspace*{-1.3cm}
\caption{The surface spectral function at $E=0$ for the $\mathcal{H}_1$ model with parameters a) $a = b = 1$, $t = -1$, $m = 0.5$, $d = 0.8$, $\alpha=\beta=0$ and b) $a = b = 1$, $t = -1.5$, $d = m = 0$, $\beta = 0.9$. To make an exact correspondence with the spinless results in Ref.~[\onlinecite{Kourtis2016}], in a) the trace is taken only over the spin-up (first and third) components of the Green's function. We clearly see that there is a single Fermi arc emerging in a), whereas there are two Fermi arcs in b). We set $U=100$.}
\label{fig:spectral_function_H1}
\end{figure}

In order to obtain the Fermi-arc surface states we need to introduce a surface into the system in such a way that the vector connecting the Weyl nodes has a nonzero projection onto it. For example, for the above model we choose to have a plane-like impurity at $y=0$, hence perpendicular to the $y$ direction.  The resulting surface Green's functions are described by the formalism in Sec.~II, where we consider an impurity $V = U  \delta(y) \mathbb{I}_4$, with $U\rightarrow \infty$ (i.e, much larger than all energy scales in the problem). Note that in Sec.~II we describe an impurity at $x=0$ and not $y=0$, however the $y=0$ formalism is obtained by simply interchanging $x$ and $y$ in the corresponding formulas. The spectral function for the surface states $A(k_x,k_z,E) = -\frac{1}{\pi} \im\{\tr[\mathcal{G}_s\left(k_x,k_z,E\right) ]\}$ is depicted in Fig.~\ref{fig:spectral_function_H1} for two chosen configurations of parameters.

\begin{figure}
\centering
\hspace*{0cm}
\vspace*{-2cm}
\hspace*{-2.5cm}
\includegraphics[width=15cm, angle =0]{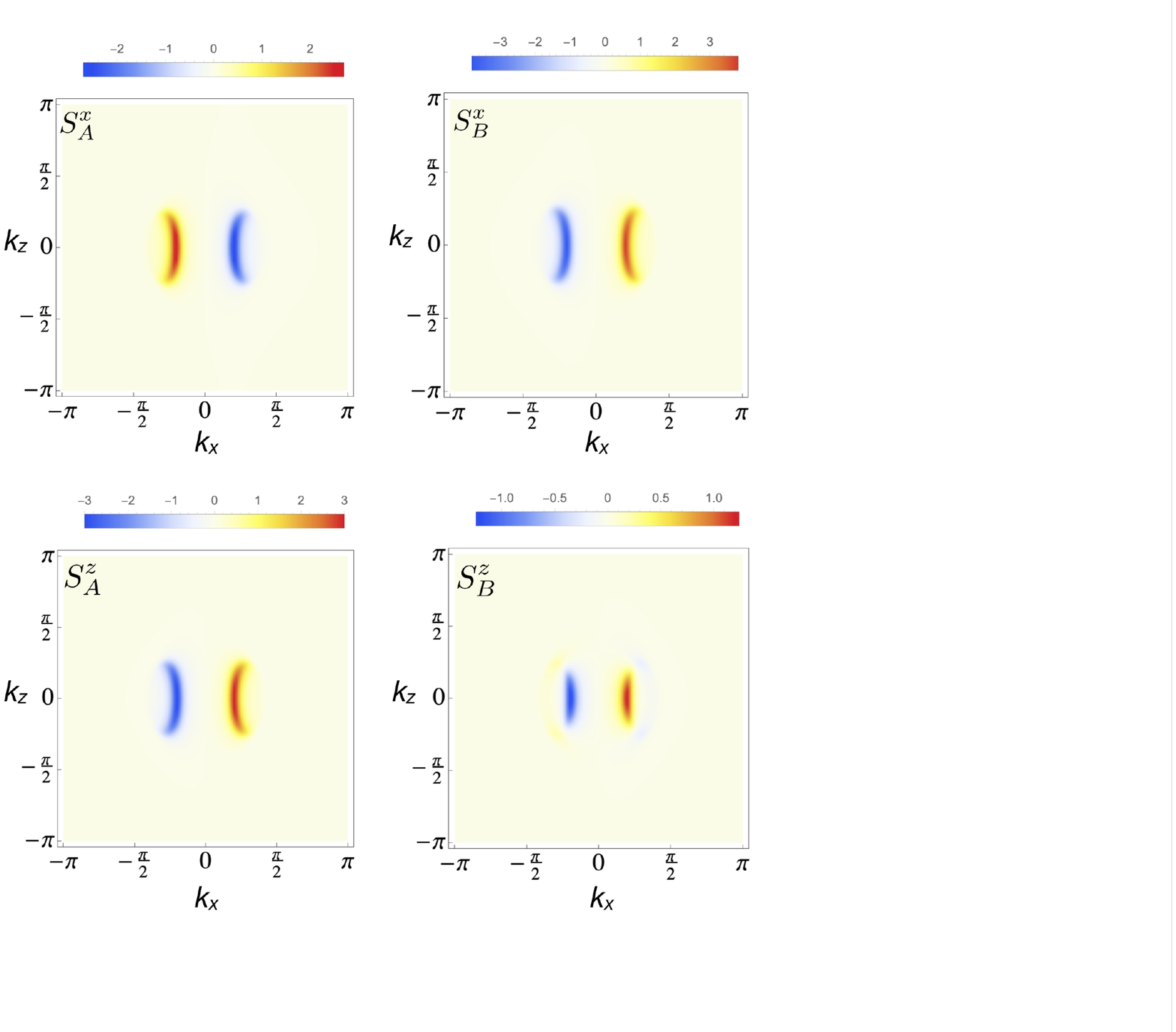}
\vspace*{0cm}
\caption{The $x$ and $z$ spin components on the A and B orbital (left and right columns, respectively) at $E=0$ for the $\mathcal{H}_1$ model with parameters $a = b = 1$, $t = 1.5$, $d = m = 0$, $\alpha=0.3$ and $\beta = 0.9$.}
\label{fig:spin_H1}
\end{figure}

We note that the Fermi-arc states calculated using our method agree exactly with those predicted in Ref.~[\onlinecite{Kourtis2016}]. Moreover, the full surface Green's function that we have obtained contains all the information required to describe these states, such as their spin and orbital distribution, the full energy dispersion, etc.. For instance, we have for the different spin and orbital components: 
\begin{eqnarray}
&&S^x_A(k_x,k_z) = -\frac{1}{\pi} \im\left[\mathcal{G}^s_{12}\left(k_x,k_z\right) +\mathcal{G}^s_{21}\left(k_x,k_z\right)\right]\nonumber \\
&&S^x_B(k_x,k_z) = -\frac{1}{\pi} \im\left[\mathcal{G}^s_{34}\left(k_x,k_z\right) +\mathcal{G}^s_{43}\left(k_x,k_z\right)\right]\nonumber \\
&&S^y_A(k_x,k_z) = -\frac{1}{\pi} \re\left[\mathcal{G}^s_{12}\left(k_x,k_z\right) -\mathcal{G}^s_{21}\left(k_x,k_z\right)\right]\nonumber \\
&&S^y_B(k_x,k_z) = -\frac{1}{\pi} \re\left[\mathcal{G}^s_{34}\left(k_x,k_z\right) -\mathcal{G}^s_{43}\left(k_x,k_z\right)\right]\nonumber \\
&&S^z_A(k_x,k_z) = -\frac{1}{\pi} \im\left[\mathcal{G}^s_{11}\left(k_x,k_z\right) -\mathcal{G}^s_{22}\left(k_x,k_z\right)\right]\nonumber \\
&&S^z_B(k_x,k_z) = -\frac{1}{\pi} \im\left[\mathcal{G}^s_{33}\left(k_x,k_z\right) -\mathcal{G}^s_{44}\left(k_x,k_z\right)\right],
\end{eqnarray} 
where we omit the energy dependence for the sake of brevity. In Fig.~\ref{fig:spin_H1} we plot the $x$ and $z$ spin components of the Fermi-arc states at zero energy, separately calculated for the A and B orbitals. The parameters chosen correspond to Fig.~\ref{fig:spectral_function_H1} b). We do not plot the $y$ component since it is zero for both orbitals.

The spins of opposite arcs are of opposite signs, as expected \cite{Kourtis2016}.

We perform a similar analysis on a different Weyl semimetal model, introduced in Ref.~[\onlinecite{Lau2017}]: 
\begin{eqnarray}
\mathcal{H}_{2}(\bs{k})&=&g_1(\bs{k}) \tau_1\sigma_3+g_2(\bs{k}) \tau_2 \sigma_0+g_3(\bs{k})\tau_3\sigma_0+d \tau_2\sigma_3
\nonumber\\&&+\beta \tau_2\sigma_2+\alpha \sin k_y \tau_1\sigma_2+\lambda \sin k_z \tau_0 \sigma_1
\end{eqnarray}

\begin{figure}
\centering
\vspace*{-0.7cm}
\hspace*{-0.5cm}
\includegraphics[width=9.6cm]{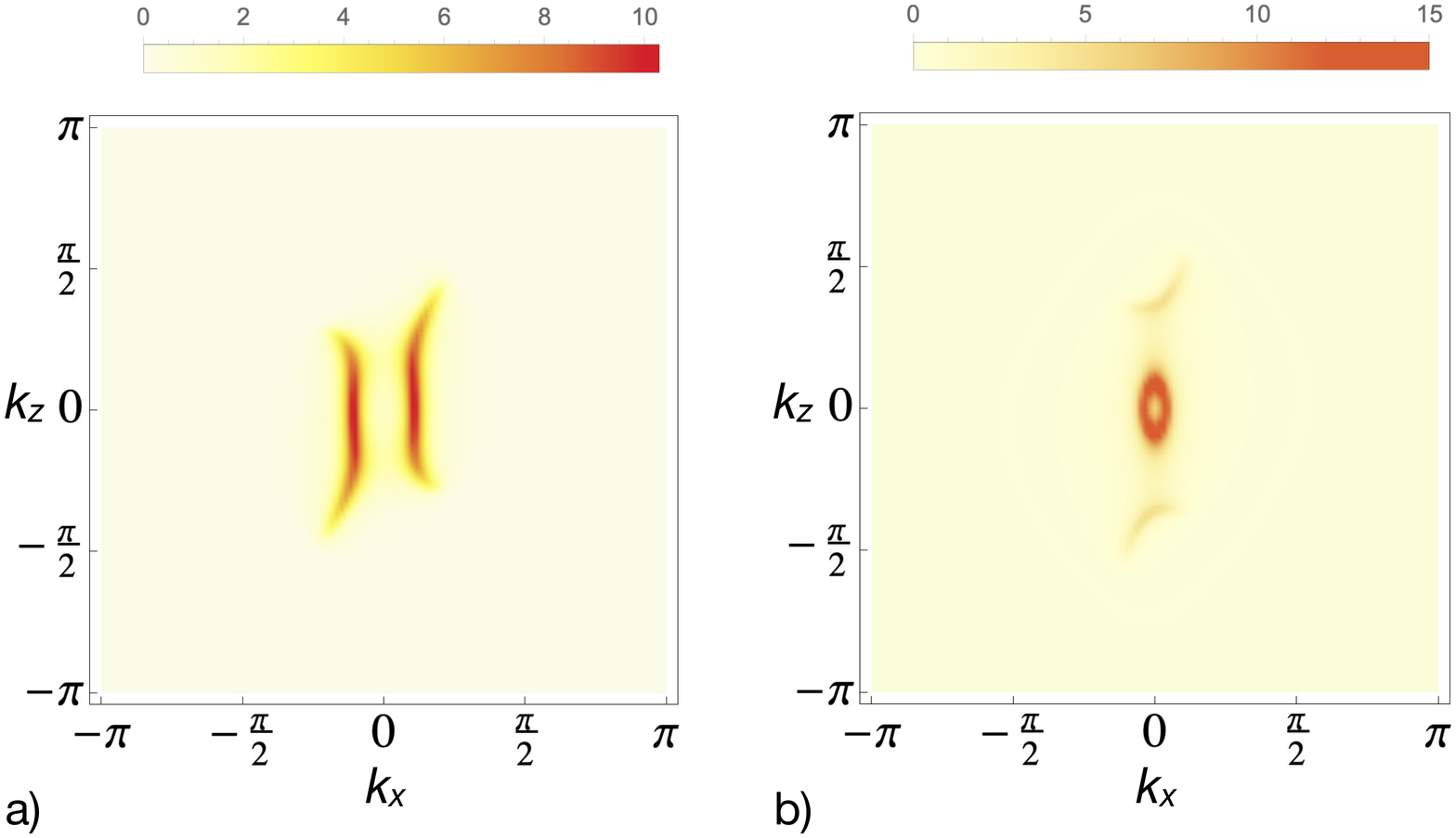}
\vspace*{-1.3cm}
\caption{The surface spectral function at $E=0$ for the $\mathcal{H}_2$ model with parameters a) $a = b = 1$, $t = -1.5$, $\lambda = 0.5$, $d = 0.1$, $\alpha=0.3$ and $\beta=0.7$ and b) $a = b = 1$, $t = -1.5$, $\lambda = 0.5$, $d = 0.1$, $\alpha=0.3$ and $\beta=0.4$.  We note the emergence of the two Fermi-arcs, as well as of the bulk electron pocket in the second configuration.}
\label{fig:spectral_function_H2}
\end{figure}

We consider the same values of parameters as those in Ref.~[\onlinecite{Lau2017}], thus we take a) $a = b = 1$, $t = -1.5$, $\lambda = 0.5$, $d = 0.1$, $\alpha=0.3$ and $\beta=0.7$ and b) $a = b = 1$, $t = -1.5$, $\lambda = 0.5$, $d = 0.1$, $\alpha=0.3$ and $\beta=0.4$. These configurations are characterized by four Weyl points, and thus one expects two Fermi arcs on the surface, however Ref.~[\onlinecite{Lau2017}] indicates the possible existence of an electron pocket coming from the bulk bands. We apply the same techniques as above, and we show in Fig.~\ref{fig:spectral_function_H2} the resulting spectral function $A(k_x,k_z,E)$ for the surface states, for the two chosen configurations of parameters.
\begin{figure}
\centering
\hspace*{0cm}
\label{fig:spin_H2}
\hspace*{0cm}
\vspace*{-1cm}
\hspace*{-1.2cm}
\includegraphics[width=15cm, angle =270]{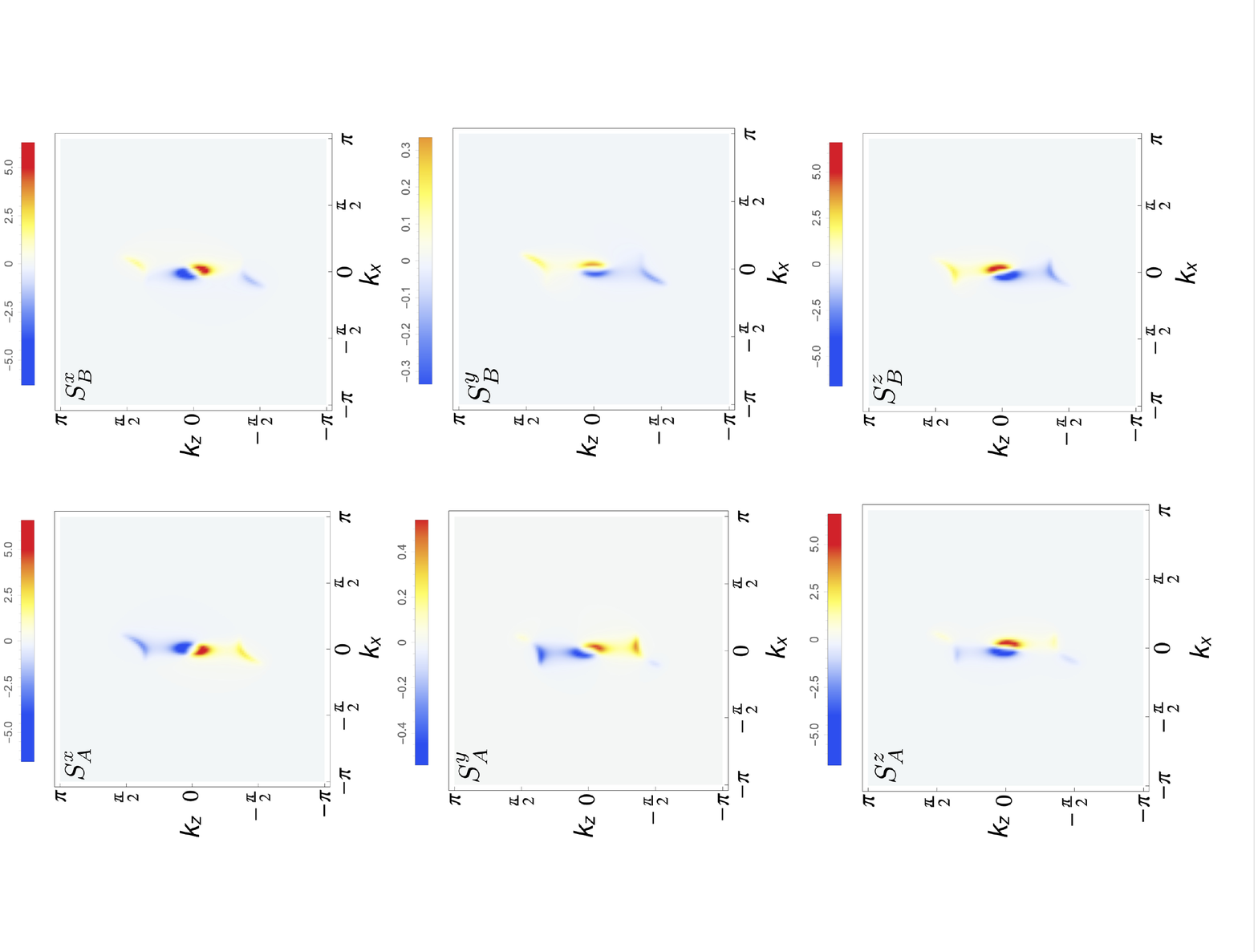}
\vspace*{0cm}
\caption{The spin components at $E=0$ computed on the A and B orbitals (left and right columns, respectively) for the $\mathcal{H}_2$ model with parameters $a = b = 1$, $t = 1.5$, $d = m = 0$, $\alpha=0.3$ and $\beta = 0.9$.}

\end{figure}

Our results agree exactly with those in Ref.~[\onlinecite{Lau2017}]. Furthermore, we compute the spin and orbital properties for this model (see Fig.~\ref{fig:spin_H2}). 

We have also checked that for the first set of parameters in Fig.~\ref{fig:spectral_function_H2} the spins of the two Fermi arcs are opposite, same as for the $\mathcal{H}_1$ model. The $\mathcal{H}_1$  and $\mathcal{H}_2$ models differ in this case mainly by a nonzero $y$ component in the $\mathcal{H}_2$ model and the asymmetry of the two Fermi arcs in $k_z$.

\section{Kane--Mele model}

\begin{figure}[ht!]
\centering
\includegraphics[width=8cm,trim={0cm 0cm 11cm 3cm},clip]{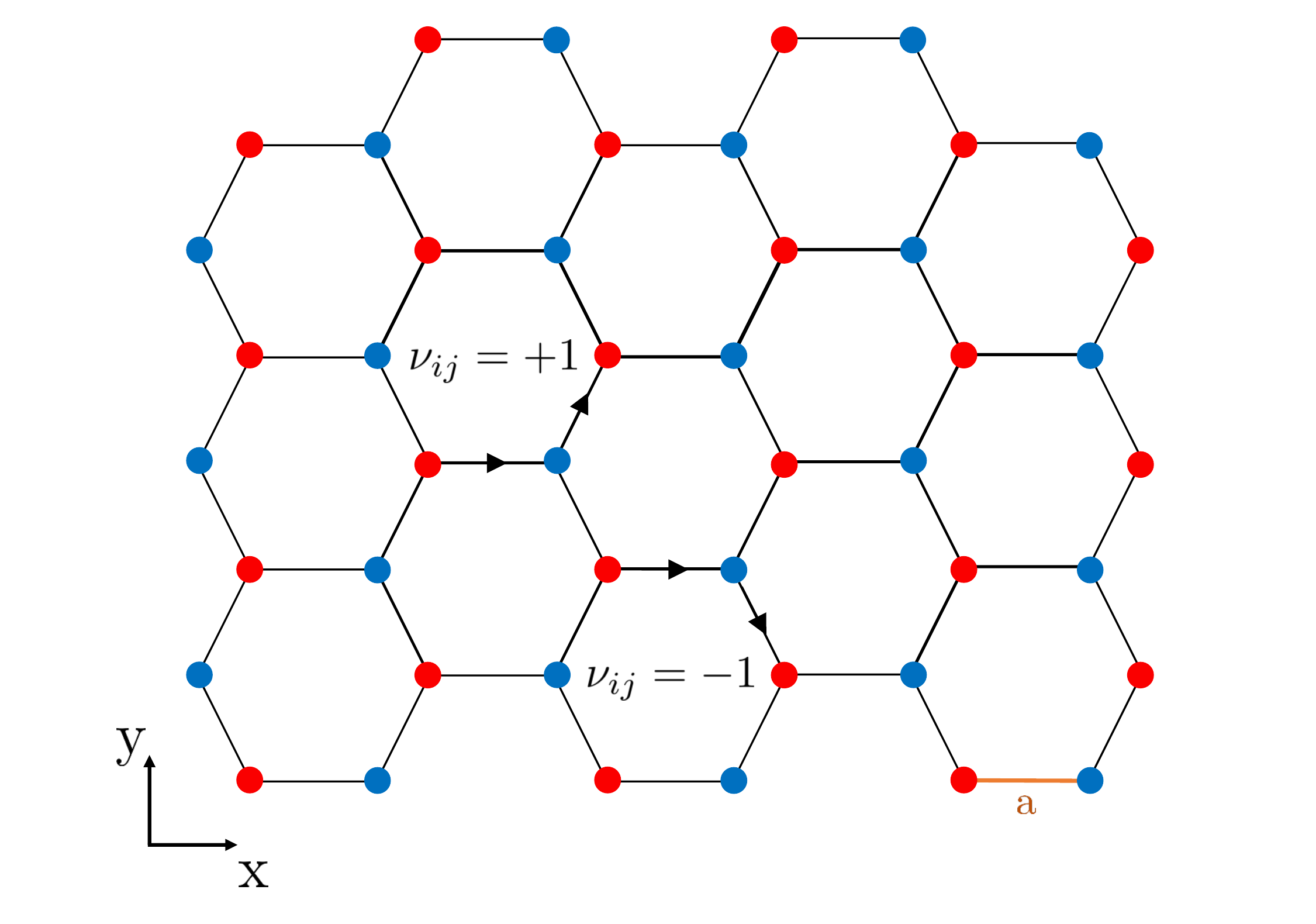}
\caption{Honeycomb lattice with the $\nu_{ij}$ convention.}
\label{fig:honeycomb}
\end{figure}

We start with the Kane--Mele model of a topological insulator on a honeycomb lattice \cite{Kane2005}. Therefore, we employ the following tight-binding model:
\begin{equation}
\label{eq:hamiltonian}
\mathcal{H}_{\mathrm{TB}} = \sum_{\langle ij \rangle, \, \alpha} t c_{i,\alpha}^\dagger c_{j,\alpha} + \sum_{\langle \langle ij \rangle \rangle, \, \alpha, \, \beta} i t_2 \nu_{ij} s_{\alpha \beta}^z c_{i,\alpha}^\dagger c_{j,\beta}
\end{equation}
where $c_{i,\alpha}^\dagger$ creates an electron on the lattice site $i$, with spin $\alpha =  \uparrow, \downarrow$. The first term in Eq.~(\ref{eq:hamiltonian}) is the standard nearest-neighbor hopping term corresponding to the tight-binding Hamiltonian of graphene, which yields a spectrum with bands touching at the Dirac points situated at the Brillouin zone corners $(\pm 4\pi/3\sqrt{3}, 0)$, $(\pm 2\pi/3\sqrt{3}, \pm 2\pi/3)$. In order to turn graphene into an insulator we add a next-nearest-neighbor term with a spin-dependent amplitude $\nu_{ij} = -\nu_{ji} = \pm 1$, defined by the orientation of the hopping direction
(see Fig.~\ref{fig:honeycomb}). The second term opens a bulk gap in the energy spectrum at the Dirac points.  

\begin{figure}[ht!]
\centering
\hspace*{-0.5cm}
\includegraphics[width=4.5cm,trim={0cm 0.5cm 0cm 0cm},clip]{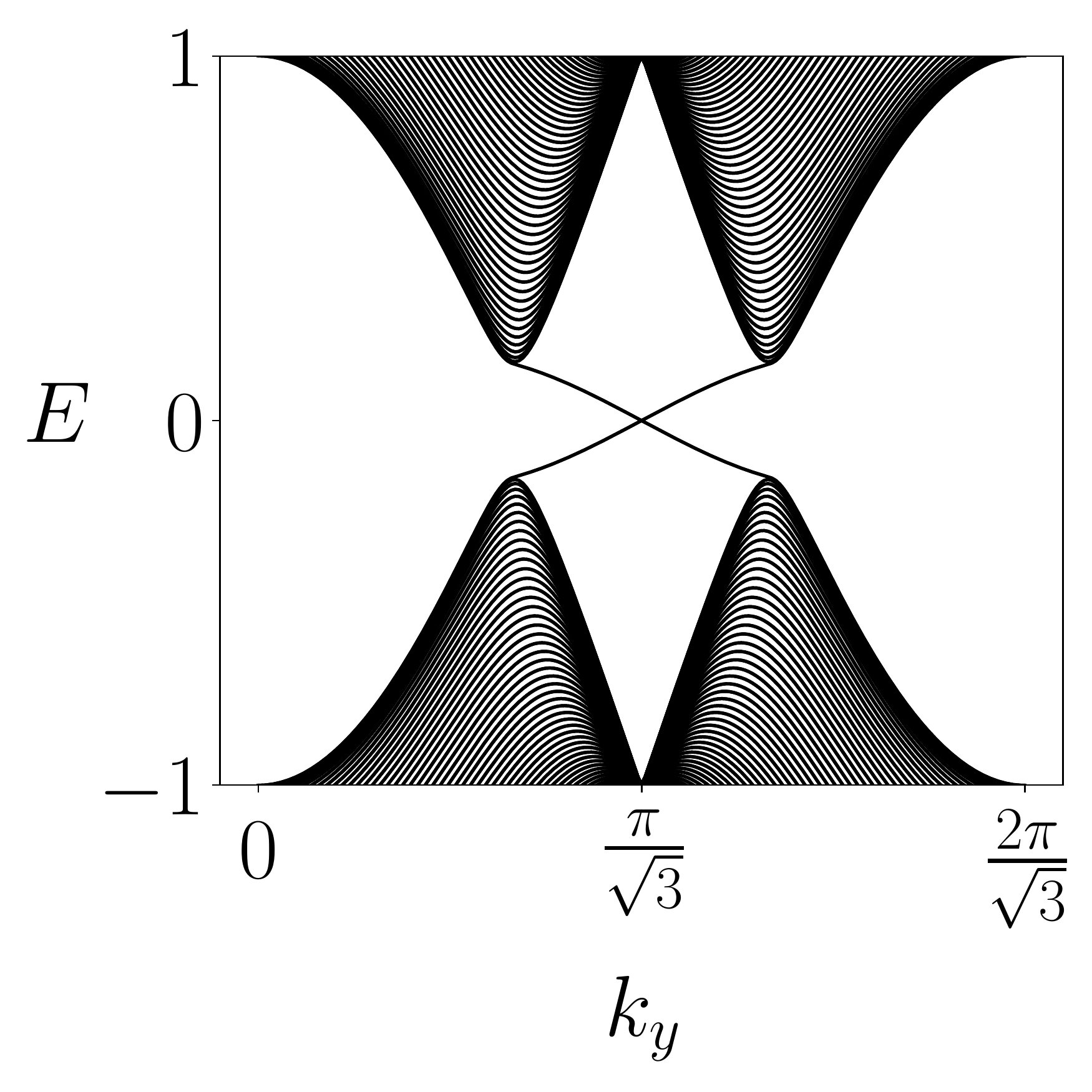}
\caption{Energy spectrum obtained by an exact diagonalization of the Hamiltonian in Eq.~(\ref{eq:hamiltonian}) defined on a strip with zigzag edges. We set $t=1$ and $t_2=0.03$. Note the formation of dispersing topological edge states in the same interval in which regular zero-energy nondispersing edge states form for a regular zigzag edge graphene nanoribbon.}
\label{fig:zigzag_spectrum}
\end{figure}

First, we obtain the boundary modes numerically by diagonalizing the tight-binding Hamiltonian in Eq.~(\ref{eq:hamiltonian}) and considering periodic boundary conditions in the $y$ direction and open boundary conditions in the $x$ direction. This corresponds to a ribbon with zigzag edges. For convenience we set the lattice spacing $a$ to unity. The corresponding energy spectrum is shown in Fig.~\ref{fig:zigzag_spectrum}. Note the formation of two subgap states (we have verified that these are actually edge states). The momentum-space dispersions for the two edge states cross at zero energy for $k_y=\pi / \sqrt{3}$.

In what follows we reproduce the formation of these edge states by considering a line impurity in an infinite system and subsequently taking the impurity potential to infinity. We can rewrite the tight-binding Hamiltonian in  Eq.~(\ref{eq:hamiltonian}) in momentum space. Thus in the basis $(c_{A\uparrow}, c_{A\downarrow}, c_{B\uparrow}, c_{B\downarrow})$, where $c_{i \sigma}$ is an electron operator with spin $\sigma=\uparrow/\downarrow$ on the sublattice $i=A/B$, the Kane-Mele Hamiltonian is expressed as:
\begin{equation}
\label{eq:hamiltonian_infinite}
\mathcal{H}_{\bs{k}} = 
\begin{pmatrix}
h_{NNN} & 0 & h_{NN} & 0 \\
0 & -h_{NNN} & 0 & h_{NN} \\
h_{NN}^* & 0 & -h_{NNN} & 0 \\
0 & h_{NN}^* & 0 & h_{NNN} 
\end{pmatrix},
\end{equation}
with 
\begin{eqnarray}
\label{eq:NN_NNN_TI}
&& h_{NN} = t \left[1 + e^{i \sqrt{3} k_y} + e^{i \frac{\sqrt{3}}{2} k_y} e^{-i \frac{3}{2} k_x} \right], {\rm and} \nonumber \\ 
&& h_{NNN} = 2 t_2 \left[2 \cos \left(\frac{3}{2}k_x\right) \sin \left(\frac{\sqrt{3}}{2} k_y\right) - \sin \left(\sqrt{3} k_y\right) \right] \nonumber
\end{eqnarray}
being the nearest-neighbor and the next-nearest-neighbor terms with amplitudes $t$ and $t_2$, respectively. Here $^*$ simply denotes the complex conjugation. 

\begin{figure}
    \centering
    \hspace*{-1cm}
    \begin{minipage}{0.24\textwidth}
        \centering
        \includegraphics[width=4.5cm,trim={0cm 0cm 0cm 0cm},clip]{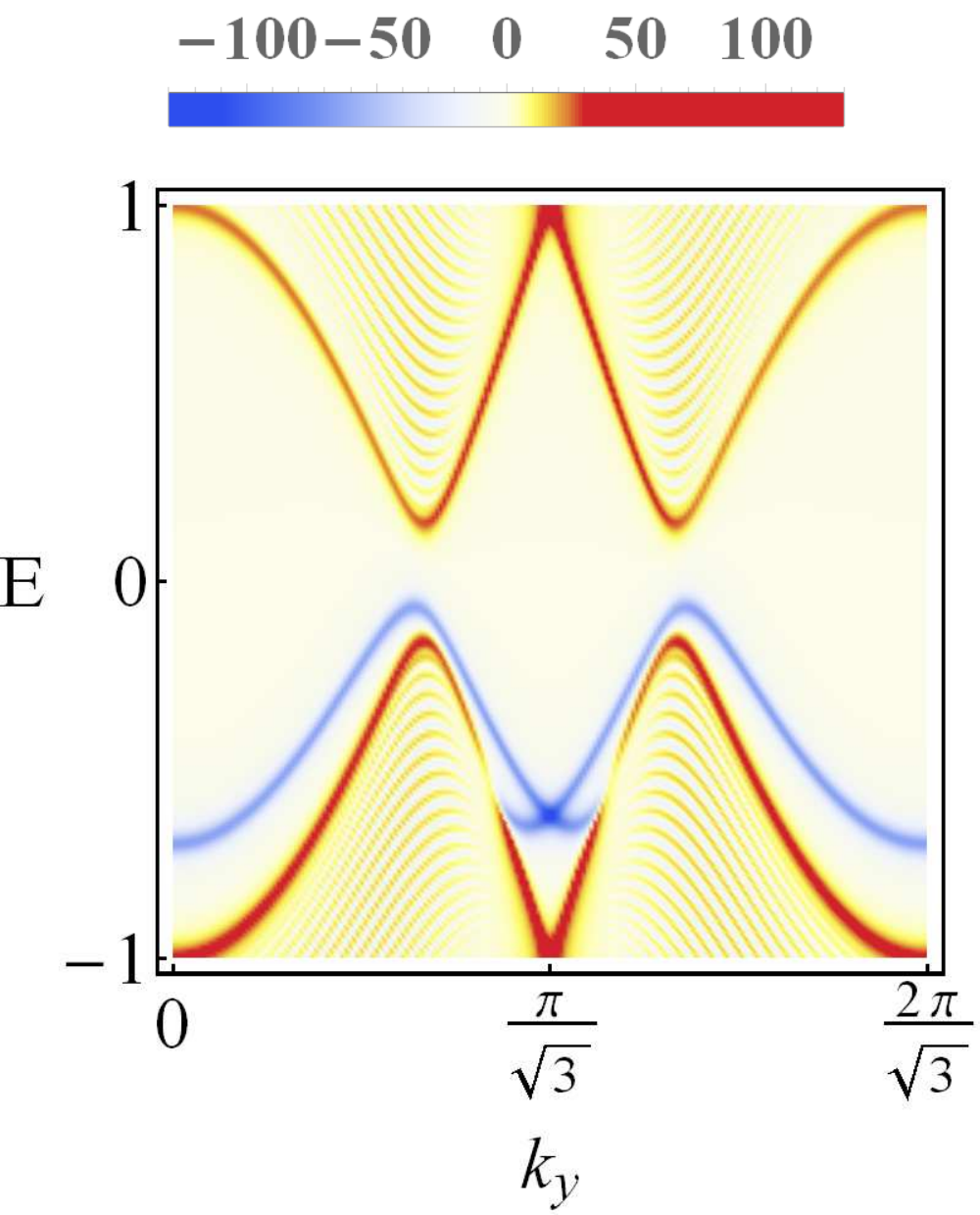}
    \end{minipage}
    \hspace*{0.3cm}
    \begin{minipage}{0.24\textwidth}
        \centering
        \includegraphics[width=4.5cm,trim={0cm 0cm 0cm 0cm},clip]{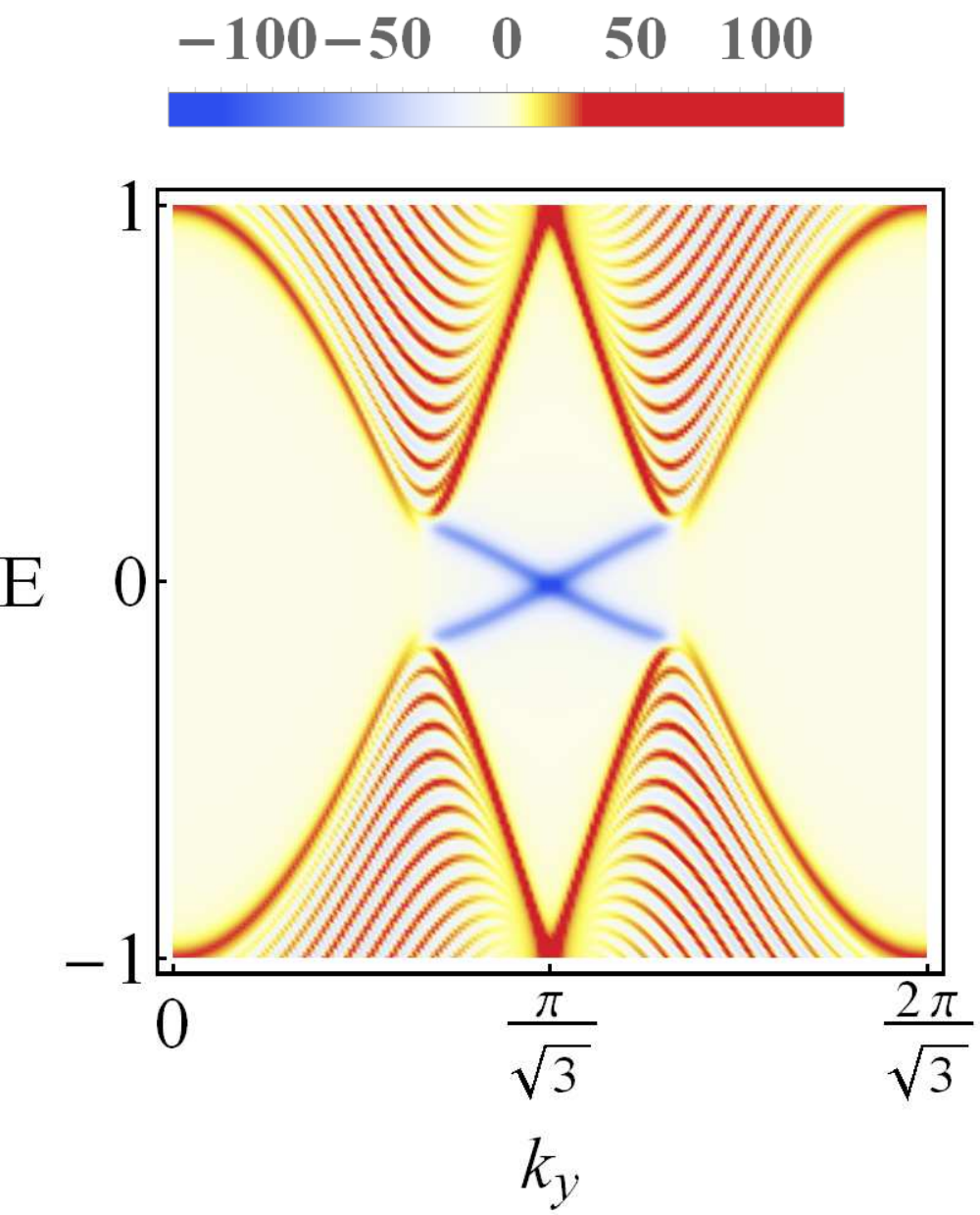}
    \end{minipage}
\caption{The average correction to the spectral function due to the line impurity, for the same energy range and momentum values as in Fig. \ref{fig:zigzag_spectrum}. Hopping amplitudes are taken to be $t=1, t_2=0.03$, and we consider $U=1$ in the left panel and $U=100$ in the right panel. Note the formation of impurity states becoming edges states at large values of the impurity potential.}
\label{fig:spectral}
\end{figure}

To reproduce the zigzag edge states, we choose an impurity potential localized on two adjacent rows of atoms corresponding to two different sublattices.
\begin{equation}
\label{eq:V_matrix_TI}
V = U
\begin{pmatrix}
1 & 0 & 0 & 0 \\
0 & 1 & 0 & 0 \\
0 & 0 & 1 & 0 \\
0 & 0 & 0 & 1 
\end{pmatrix},
\end{equation}
(see Appendix \ref{App:KMimpurity} for more details). 

In order to visualize the impurity-induced states, in Fig.~\ref{fig:spectral}  we plot the average correction to the spectral function due to the impurity, as defined in Eq.~(\ref{def}),
\begin{equation}
\delta N(k_y,E)=\int \frac{dk_x}{L_k} \delta A(k_x,k_y,E),
\label{adef}
\end{equation}
for the same range of $k_y$ as the one considered in Fig. \ref{fig:zigzag_spectrum}. 

For a weak impurity, i.e., $U=t=1$, the impurity states appear as a distinct band at energies concentrated mostly outside of the gap. We expect that the impurity bound states will evolve into edge states and acquire the same properties (i.e., the same momentum dispersion) as the edge states derived previously using numerical methods, when the impurity strength $U$ goes to infinity. Indeed, for a stronger impurity potential $U=100$, the impurity-induced spectral function exhibits subgap states with the same dispersion as the ones derived via exact diagonalization and depicted in Fig.~\ref{fig:zigzag_spectrum}. The agreement between the two methods is remarkable, confirming the validity of our analytical approach towards finding the edge states of a simple topological insulator system.

While here we consider only zigzag edges, in Appendix \ref{App:KMimpurity} we have also considered the case of an impurity localized only on one row of atoms, which splits the systems into two subsystems with different edges, one with a zigzag edge, and one with a bearded edge. We expect that we will recover two distinct sets of edge states, and in Appendix \ref{App:KMimpurity} we show that this is indeed the case.
\\

\section{Chern insulator}

Below we consider the simplest lattice model defining a Chern insulator 
\begin{align}
\nonumber \mathcal{H}_{\bs{k}} &= t \sin k_x \sigma_x +t \sin k_y \sigma_y \\
& + B(2-M-\cos k_x -\cos k_y) \sigma_z
\label{eq:HChern}
\end{align}
where we set the lattice constant to unity and $t=1$. Here $\bs{k} \equiv (k_x,\,k_y)$ and $\bs{\sigma} = (\sigma_x,\,\sigma_y,\,\sigma_z)$ are the Pauli matrices. The subspace in which they act may be very general and depends on the given model, for example in a lattice model with two orbitals per site, $s$ and $p$, the $\bs{\sigma}$ matrices act in the orbital subspace.  The above model yields topologically nontrivial phases for $M \in (0,\,2) \cup (2,\,4)$ (see Ref.~[\onlinecite{Bernevig2013}]).  

In what follows we introduce a line-like impurity at $x=0$  described by the potential $V_{\mathrm{imp}}(x) = V \delta(x) \mathbb{I}$, with $V \to \infty$ and $\mathbb{I}$ is the $2\times 2$ identity matrix. In this limit the $T$-matrix can be written as:
\begin{align}
\nonumber T(k_y, i \omega_n) &= \negthickspace\lim\limits_{V \to \infty}\left[\mathbb{I} - V \negthickspace \int\limits_{-\pi}^{\pi}\negthickspace\frac{dk_x}{2\pi} G_0(k_x,k_y,i\omega_n)\right]^{-1} \negthickspace V  \\
&= - \left[\int\limits_{-\pi}^{\pi} \frac{dk_x}{2\pi} G_0(k_x,k_y,i\omega_n)\right]^{-1}
\label{TmatrixChern}
\end{align}
We compute the integral in Eq.~(\ref{TmatrixChern}), setting $B = M = 1$ for the sake of simplicity. We note that the calculation can be performed for arbitrary values of $B$ and $M$. Thus, we have 
\begin{widetext}
\begin{align}
\nonumber \int\limits_{-\pi}^{\pi} \frac{dk_x}{2\pi} G_0(k_x,k_y,i\omega_n) = \frac{1}{4\sin^2\frac{k_y}{2}}\frac{1}{\sqrt{(i\omega_n)^2-1}\sqrt{(i\omega_n)^2-5+4\cos k_y}} \times \phantom{aaaaaaaaaaaaaaaaaaaaaaaaaaaaaaaa}\\ 
\times\left\{
	\left[ (i\omega_n)^2 -2\cos k_y + \cos 2k_y - \sqrt{(i\omega_n)^2-1}\sqrt{(i\omega_n)^2-5+4\cos k_y} \right]\sigma_z + 4\sin^2\frac{k_y}{2}( i\omega_n\sigma_0 + \sin k_y \sigma_y)
\right\}.
\label{IntOfG0Chern1}
\end{align}
\end{widetext}
Plugging Eq.~(\ref{IntOfG0Chern1}) into Eq.~(\ref{TmatrixChern}) we obtain the $T$-matrix, which in turn defines the perturbed GF given by Eq.~(\ref{eq:green_perturbed}).  Note that the poles of the $T$-matrix obtained by taking $i \omega_n\rightarrow E + i \delta$, with $\delta \to +0$, are given by $E = \pm \sin k_y$ for $k_y \in \left[-\pi/2,\,\pi/2 \right]$, corresponding to two chiral edge modes (cf. Ref.~[\onlinecite{Konig2008}]). For the sake of brevity, we leave the derivation of the poles of the $T$-matrix to Appendix \ref{App:TmatrixPoles}.

\begin{figure}
	\centering
	\hspace*{0cm}
	\includegraphics[width=5cm,trim={0cm 0cm 0cm 0cm},clip]{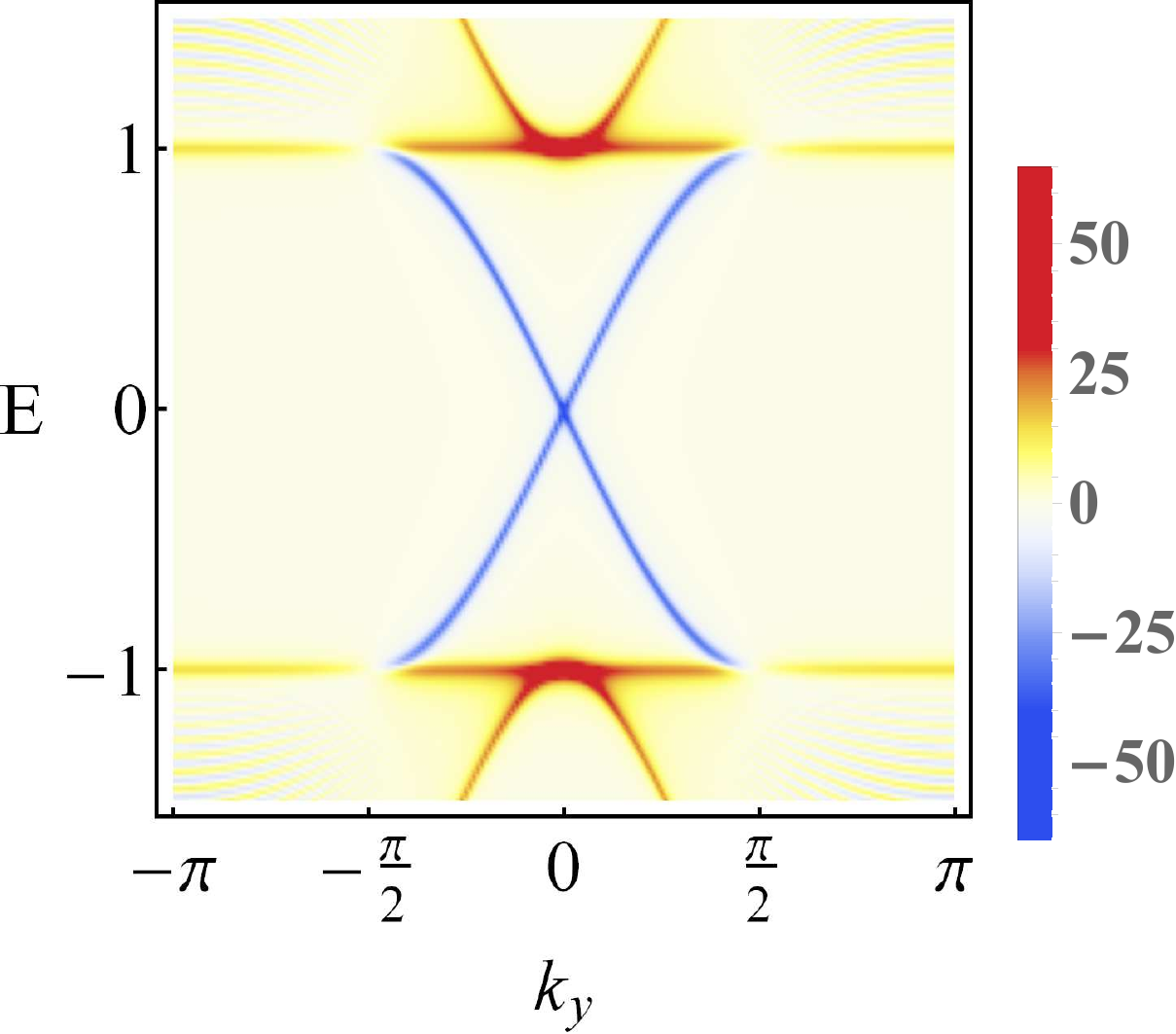}
	\caption{The average correction to the spectral function of the Chern insulator with a line-like impurity potential. The bulk bands are visible at $E=\pm 1$, whereas the edge modes disperse as $E = \pm \sin k_y$, when $k_y \in \left[-\pi/2,\,\pi/2 \right]$. We set $B=M=t=1$ and $V=100$.}
	\label{fig:ChernES}
\end{figure}

To verify our findings, in Fig.~\ref{fig:ChernES} we plot the average correction to the spectral function defined in Eq.~(\ref{def}) as a function of $E$ and $k_y$. As expected, on one hand we can see the bulk states, originating from the poles of the bare Green's function i.e., the eigenvalues of the Hamiltonian in Eq.~(\ref{eq:HChern}) which for the values considered here, $B=M=t=1$ and $k_x=0$ correspond to $E = \pm 1$. More importantly we can identify also the two counter-propagating chiral edge modes of the Chern insulator crossing at $k_y = 0$, whose dependence on $k_y$ is consistent with the fully analytical form above. This demonstrates the strength of our approach to recover fully analytical results for the edge states of certain models for which the unperturbed Green's function in the real space can be obtained in an analytical closed form.

\section{Conclusions}

We have generalized the technique to obtain boundary modes, introduced in Ref.~[\onlinecite{Kaladzhyan2018}], to calculate the surface Green's functions  of an arbitrary three-dimensional system, and we have applied it to calculate the surface Green's functions for Weyl semimetals and recovered the corresponding Fermi-arc states. We have also shown that the technique in Ref.~[\onlinecite{Kaladzhyan2018}] can be applied to other topological systems, such as topological insulators. Furthermore, we have demonstrated that it functions also for systems with more than one sublattice, and that this method can be easily employed to study boundary modes in lower dimensions. In particular, using line-like impurities, we have applied the formalism to derive the helical edge states of the Kane--Mele model and the chiral edge states of a Chern insulator. For the latter we have shown that using this formalism a full analytical form can be obtained for the $T$-matrix, and analytical expressions for the energies of the edge states can be recovered. \\

\begin{acknowledgments}
V.K. would like to acknowledge the ERC Starting Grant No. 679722 and the Roland Gustafsson foundation for theoretical physics.
\end{acknowledgments}

\bibliography{biblio_Tmatrix}

\widetext
\appendix

\section{Derivation of the T-matrix momentum limits of integration for a honeycomb lattice}\label{FT}
While for a square or cubic lattice there is no subtlety concerning the integration limits, for the honeycomb lattice this is much more subtle. We will start with writing down the form of the contribution of the impurity potential to the Hamiltonian, such as it has to be written in the continuum. We will start with a row of impurities localized on A atoms, but the conclusions do not depend on the type of impurity we consider:
\be
\delta H_{\mathrm{imp}}=\int d{\bf r}\, V({\bf r})
\ee
where the integral is performed over the entire space, with

\begin{gather}
V({\bf r}) =U \rho({\bf r}), \nonumber \\
 \rho({\bf r}) =\sum_j\delta({\bf r}-{\bf{R}}^{A}_j) a^\dagger_j a_j
\end{gather}

where the sum is performed over a row of lattice unit cells, each unit cell is denoted by the index $j$ (see Eq.~(24) in Ref.~[\onlinecite{Bena2009NJP}]) and $R_j=j a \sqrt{3}$ (see Fig.~9), with the lattice constant $a$ having been set to 1. The $a^\dagger_j$ and $a_j$ operators describe the formation and annihilation of electrons at site $j$, they no longer live in the continuum but on the lattice and as such they are defined as (see Eq.~(21) in Ref.~[\onlinecite{Bena2009NJP}]):
\be
a_j=\int_{{\bf k} \in BZ} d{\bf k}\, e^{-i {\bf k}\cdot {\bf{R}}^{A}_j} a_{\bf k}
\ee
where $\int_{{\bf k} \in BZ} \equiv \int_{BZ} \frac{d^2 k}{S_{BZ}}$ with $S_{BZ}=\frac{8 \pi^2}{3 \sqrt{3}}$.
In order to use the momentum space T-matrix formalism we need to write $\delta H_{\mathrm{imp}}$ in momentum space:

\begin{align}
\delta H_{\mathrm{imp}}&=U\int d{\bf r} \sum_j\delta({\bf r}-{\bf{R}}^{A}_j) \int_{{\bf k} \in BZ} \int_{{\bf k'} \in BZ} d{\bf k} d{\bf k'}\, e^{i ({\bf k}-{\bf k'})\cdot {\bf{R}}^{A}_j}  a^{\dagger}_{\bf k} a_{\bf k'} \nonumber \\ 
&=U\sum_j \int_{{\bf k} \in BZ} \int_{{\bf k'} \in BZ} d{\bf k} d{\bf k'}\, e^{i ({\bf k}-{\bf k'})\cdot {\bf{R}}^{A}_j}  a^{\dagger}_{\bf k} a_{\bf k'} \nonumber \\
&=U\sum_j \int_{{\bf k} \in BZ} dk_x dk_y \int_{{\bf k'} \in BZ} dk'_x dk'_y\, e^{i (k_y-k'_y) (\sqrt{3} j)} a^{\dagger}_{\bf k} a_{\bf k'} \nonumber\\
&=U \frac{2\pi}{\sqrt{3}} \sum_{n} \int_{{\bf k} \in BZ} dk_x dk_y \int_{{\bf k'} \in BZ} dk'_x dk'_y\, \delta(k_y-k'_y+n \frac{2\pi}{\sqrt{3}}) a^{\dagger}_{\bf k} a_{\bf k'}
\end{align}

Since both $k$ and $k'$ are in the first BZ the only possibilities for $n$ are $0,1$ and $-1$. It appears that implementing this constraint is quite subtle, however things get much simpler if we deform the first BZ and instead we consider a rectangle with $-\frac{2\pi}{3\sqrt{3}}<k_y<\frac{4\pi}{3\sqrt{3}}$ and $-2\pi/3<k_x<2\pi/3$. This is allowed since $a_{k+{\bf Q}_{\mu \nu}}=a_k$, where $ {\bf Q}_{\mu \nu}$ are all the reciprocal basis vectors (see Ref.~[\onlinecite{Bena2009PRB}]; in the tight-binding basis considered here the same relation is valid also for the $B$ atoms \cite{Bena2009NJP}). Under this construction it is clear that the only possible solution for $n$ is $n=0$, and thus we have
\be
\delta H_{\mathrm{imp}}=U  \int_{-\frac{2 \pi}{3\sqrt{3}}}^{\frac{4 \pi}{3\sqrt{3}}} \frac{dk_y}{2\pi/\sqrt{3}}\int_{-2\pi/3}^{2\pi/3} \frac{dk_x}{L_k} \int_{-2\pi/3}^{2\pi/3} \frac{dk'_x}{L_k} a^\dagger_{k_x,k_y} a_{k'_x,k_y} 
\ee
where $L_k=4\pi/3$.
\newpage
\widetext

\section{Significance of a position-averaged spectral function}\label{kxintegral}
In 2D, in the presence of a line impurity proportional to $\delta(x)$, the correction to the number of available electronic states at position $x$ and having momentum $k_y$  is given by:
\begin{align}
\delta \rho(x,k_y,E) = -\frac{1}{\pi} \im \tr \delta\mathcal{G}(x,x;k_y;E)
\end{align}
Note that since the spatial translational invariance along $y$ is not broken $k_y$ is still a good quantum number.

Averaging this over the direction perpendicular to the impurity we obtain:
\begin{align}
\nonumber \delta N(k_y,E) &\equiv \int dx\, \delta \rho(x,k_y,E) = -\frac{1}{\pi} \im \tr \int dx\,\delta\mathcal{G}(x,x;k_y;E) = -\frac{1}{\pi} \im \tr \int dx\, \mathcal{G}_0(x,k_y,E) T(k_y,E) \mathcal{G}_0(-x,k_y,E) = \\
\nonumber &= -\frac{1}{\pi} \im \tr \int dx\, \int \frac{dk_x}{L_k}\frac{dk_x'}{L_k} e^{i k_x x} e^{-i k_x' x}\, \mathcal{G}_0(k_x,k_y,E) T(k_y,E) \mathcal{G}_0(k_x',k_y,E) =\\
&= -\frac{1}{\pi} \im \tr \int \frac{dk_x}{L_k}\mathcal{G}_0(k_x,k_y,E) T(k_y,E) \mathcal{G}_0(k_x,k_y,E) \equiv \int \frac{dk_x}{L_k} \delta A(k_x,k_y,E),
\end{align}
where $L_k$ and the limits of integration are $-\pi$ to $\pi$ with $L_k=2\pi$ for a square lattice and $-2\pi/3$ to $2\pi/3$ with $L_k=4\pi/3$ for a honeycomb one. Also
\begin{equation}
\label{eqApp:spectral_function}
\delta A(k_x,k_y,E) \equiv -\frac{1}{\pi} \im \tr \delta\mathcal{G}(k_x,k_y;k_x,k_y;E) = -\frac{1}{\pi} \im \tr \mathcal{G}_0(k_x,k_y,E) T(k_y,E) \mathcal{G}_0(k_x,k_y,E)
\end{equation} 
is the correction to the perturbed spectral function in momentum space.

\section{\boldmath{$T$}-matrix poles for the Chern insulator}\label{App:TmatrixPoles}

In order to calculate the energies of the bound states, here we calculate analytically the poles of the $T$-matrix defined by Eqs.~(\ref{TmatrixChern}-\ref{IntOfG0Chern1}). The latter can be found from the trace of the $T$-matrix given by
\begin{align}
\tr T(k_y, E+i\delta) = -\frac{(E+i\delta)\left[(E+i\delta)^2 -2 \cos k_y + \cos 2k_y + \sqrt{(E+i\delta)^2-1}\sqrt{(E+i\delta)^2 - 5 + 4\cos k_y} \right]}{(E+i\delta)^2 - \sin^2 k_y},
\label{App:eq:Tmatrixtrace}
\end{align}
where we replaced $i\omega_n \to E+i\delta$, with $\delta \to +0$. We obtain straightforwardly the zeros of the denominator, namely, $E = \pm \sin k_y$. However, to make sure that the latter are poles, we need to verify that they are not zeros of the numerator. The trivial zero of the numerator is $E=0$, we discard it below. Thus, we need to analyze the zeros of the expression in the square brackets:
\begin{align}
(E+i\delta)^2 -2 \cos k_y + \cos 2k_y + \sqrt{(E+i\delta)^2-1}\sqrt{(E+i\delta)^2 - 5 + 4\cos k_y} = 0
\end{align}
We represent the complex numbers under the square roots in the trigonometric form, and applying the limit $\delta \to +0$ we get:
\begin{align}
E^2 -2 \cos k_y + \cos 2k_y + \sqrt{\left|E^2-1\right|}\, e^{i\phi_1(E)} \; \sqrt{\left|E^2 - 5 + 4\cos k_y\right|}\,e^{i\phi_2(k_y,E)} = 0,
\label{App:eq:energies}
\end{align}
where we defined
$
\phi_1(E) = \frac{\pi}{2} \sgn E\, \Theta\left(1-E^2\right),\; \phi_2(k_y,E) = \frac{\pi}{2} \sgn E\, \Theta\left(5 - 4\cos k_y - E^2\right).
$
Since we are searching for subgap solutions, i.e., $|E| < 1$, then $E^2 < 5 - 4\cos k_y\; \forall k_y \in \left[-\pi,\,\pi \right]$. Thus, we have $\phi_1(E) + \phi_2(k_y,E) = \pi \sgn E$ and, therefore, $e^{i \left[\phi_1(E) + \phi_2(k_y,E) \right]} =  e^{i \pi \sgn E} = -1$. Eq.~(\ref{App:eq:energies}) then becomes
\begin{align}
E^2 -2 \cos k_y + \cos 2k_y - \sqrt{1 - E^2} \sqrt{5 - 4\cos k_y - E^2} = 0.
\end{align}
The equation above is equivalent to solving the system:
\begin{align}
\left(E^2 - \sin^2 k_y \right) \sin^4 \frac{k_y}{2} &= 0 \\
E^2 -2 \cos k_y + \cos 2k_y &> 0
\end{align}
When $k_y = 0$, we get $E = \pm 1$ at the edge of the gap, therefore, $E = \pm \sin k_y$ from the first equation. The second equation then yields:
\begin{align}
\sin^2 k_y - \cos k_y + \cos 2k_y > 0 \;\Rightarrow\; |k_y| \in \Big(\frac{\pi}{2},\,\pi\Big].   
\end{align}
Thus, the numerator of Eq.~(\ref{App:eq:Tmatrixtrace}) has zeros at $E=0$, and $E = \pm \sin k_y$ when $|k_y| \in (\pi/2,\,\pi]$, and therefore, we conclude that the trace of the $T$-matrix has poles at $E = \pm \sin k_y$ only when $|k_y| \leqslant \pi/2$. This means that the edge modes exist only for $k_y$ lying in the interval $|k_y| \leqslant \pi/2$, and their dispersion is given by $E = \pm \sin k_y$. 

\widetext

\section{Impurity potentials for the Kane-Mele model}\label{App:KMimpurity}

\begin{figure}[ht!]
    \centering
    \hspace*{-1cm}
    \begin{minipage}{0.33\textwidth}
        \centering
        \includegraphics[width=8.5cm,trim={0cm 0cm 0cm 0cm},clip]{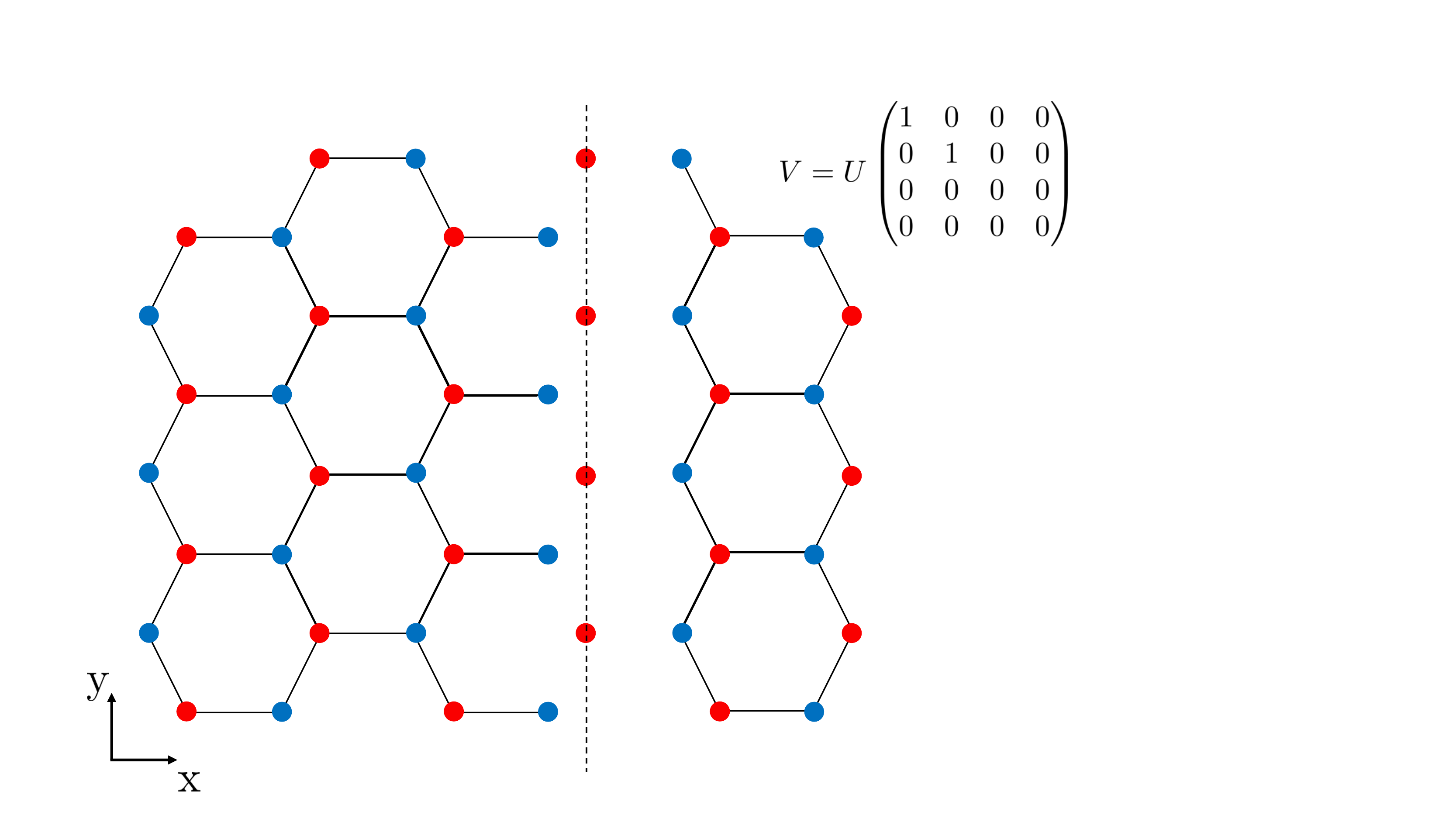}
    \end{minipage}
    \hspace*{0.3cm}
    \begin{minipage}{0.33\textwidth}
        \centering
        \includegraphics[width=8.5cm,trim={0cm 0cm 0cm 0cm},clip]{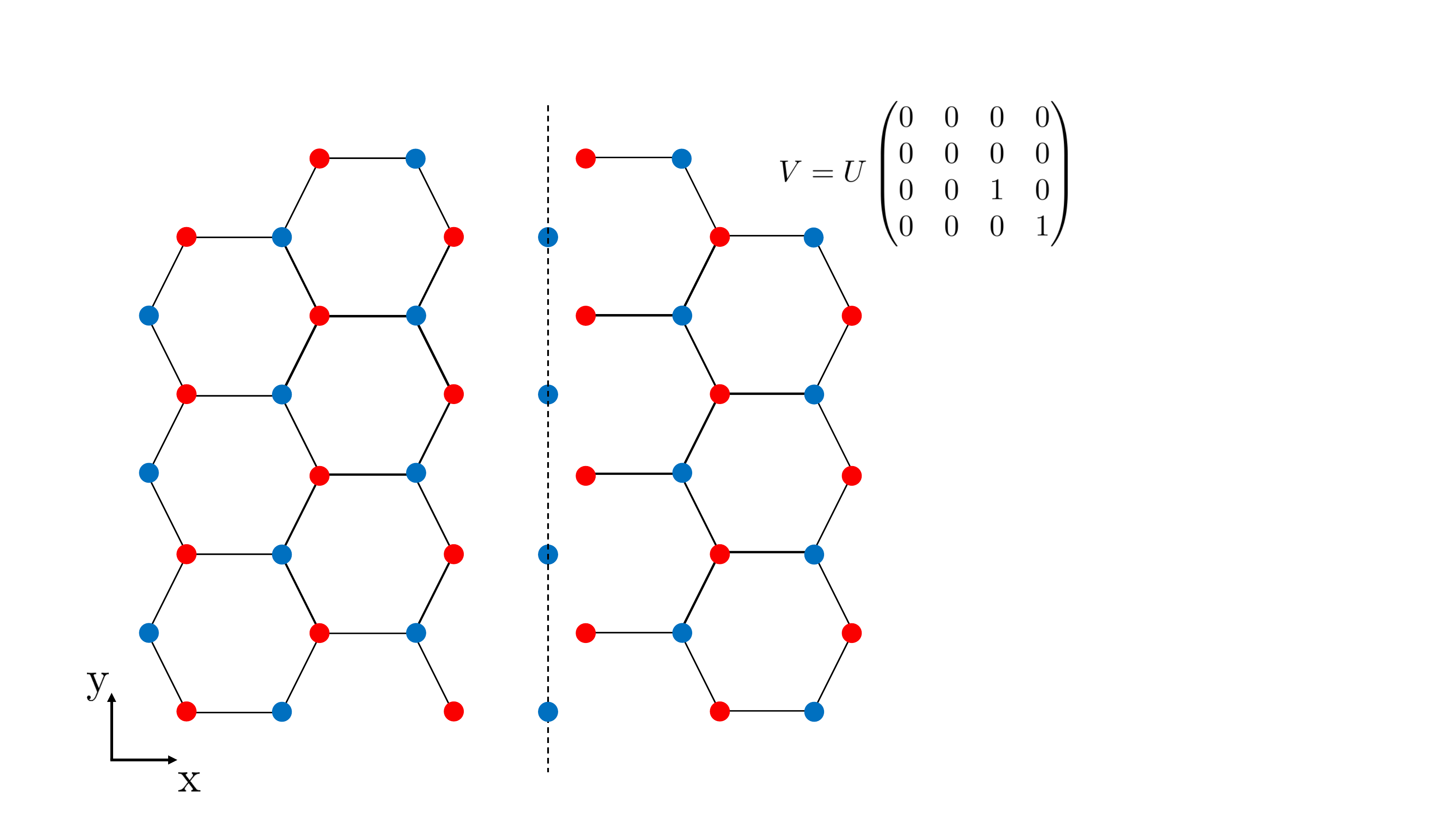}
    \end{minipage}
    \hspace*{0.3cm}
    \begin{minipage}{0.33\textwidth}
        \centering
        \includegraphics[width=8.5cm,trim={0cm 0cm 0cm 0cm},clip]{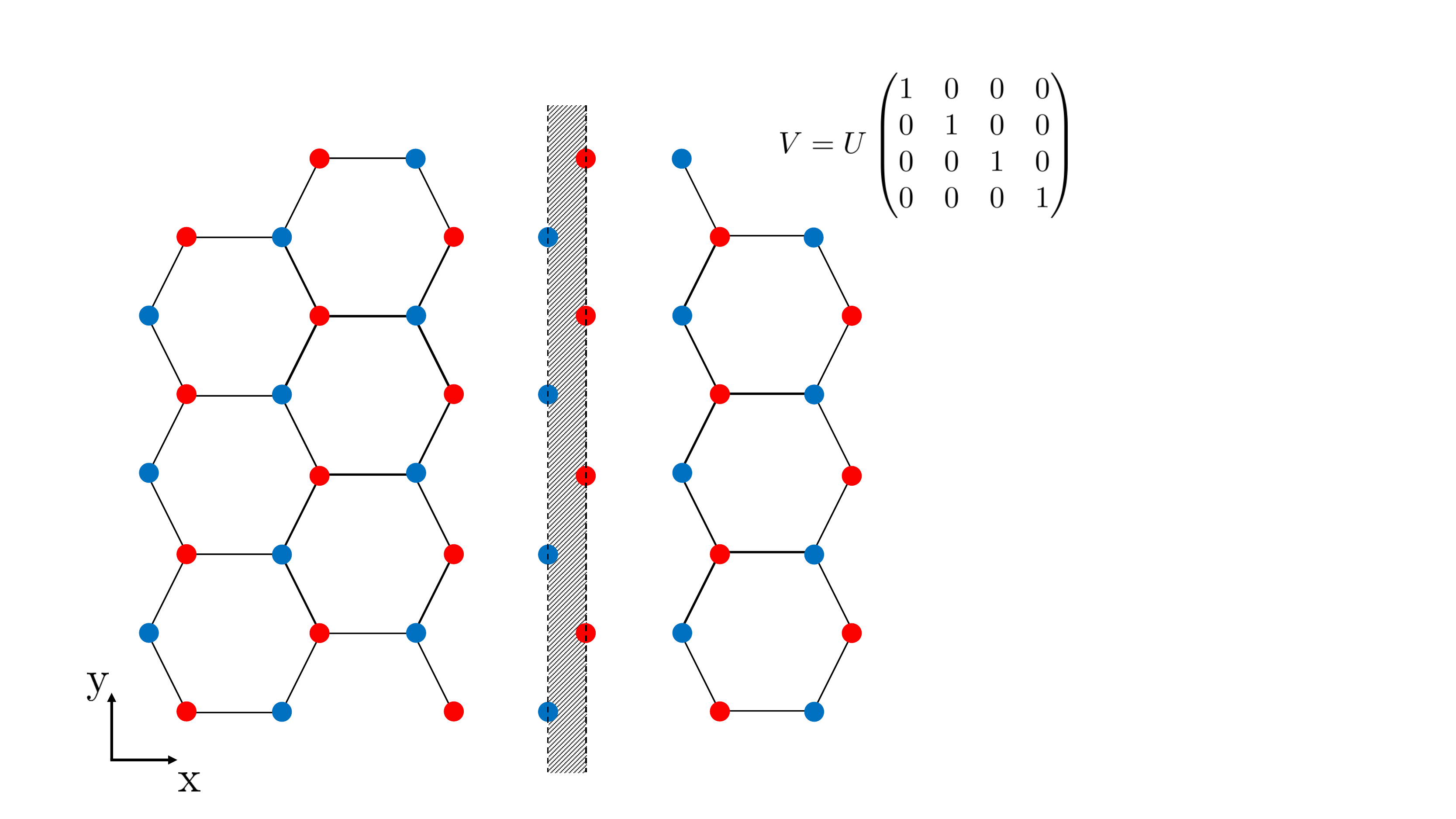}
    \end{minipage}
\caption{Three different vertical impurity lines on a honeycomb lattice. From left to right: the impurity is localized on sublattice $A$, sublattice $B$ or on entire unit cells ($A + B$). In each case, the impurity creates a ``wall" in the system (dashed lines or shaded area) along with two boundaries. If the impurity is localized on a single sublattice ($A$ or $B$), it creates one zigzag edge and one bearded edge. If it is localized on entire unit cells it will create two zigzag edges. The matrix representation in the insets is given in the basis $(c_{A\uparrow}, c_{A\downarrow}, c_{B\uparrow}, c_{B\downarrow})$.}
\label{fig:KMimpurities}
\end{figure}

While working with the Kane-Mele model, we considered three different delta-function impurities which we illustrate in Fig. \ref{fig:KMimpurities} with the associated matrix representation. We see that in order to reproduce the zigzag edge states of the Kane-Mele model we must introduce an impurity on both $A$ and $B$ sites (right panel of Fig. \ref{fig:KMimpurities}). If the impurity is localized only on the $A$ or $B$ sites, it will create both a zigzag edge state and a bearded edge state. 

\begin{figure}[ht!]
    \centering
    \hspace*{-1cm}
    \begin{minipage}{0.49\textwidth}
        \centering
        \includegraphics[width=4.5cm,trim={0cm 0cm 0cm 0cm},clip]{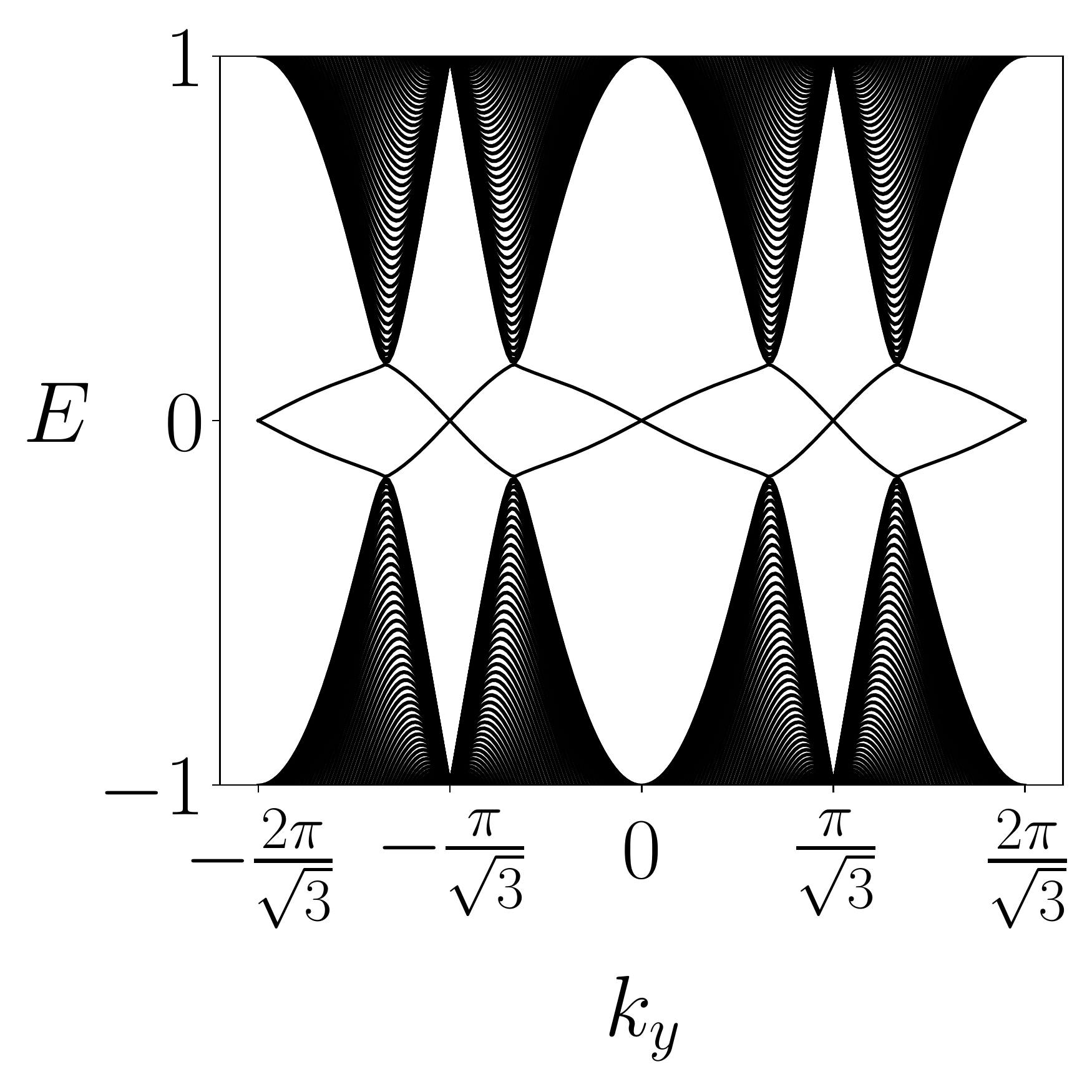}
    \end{minipage}
    \hspace*{0.3cm}
    \begin{minipage}{0.49\textwidth}
        \centering
        \includegraphics[width=4.7cm,trim={0cm 0cm 0cm 0cm},clip]{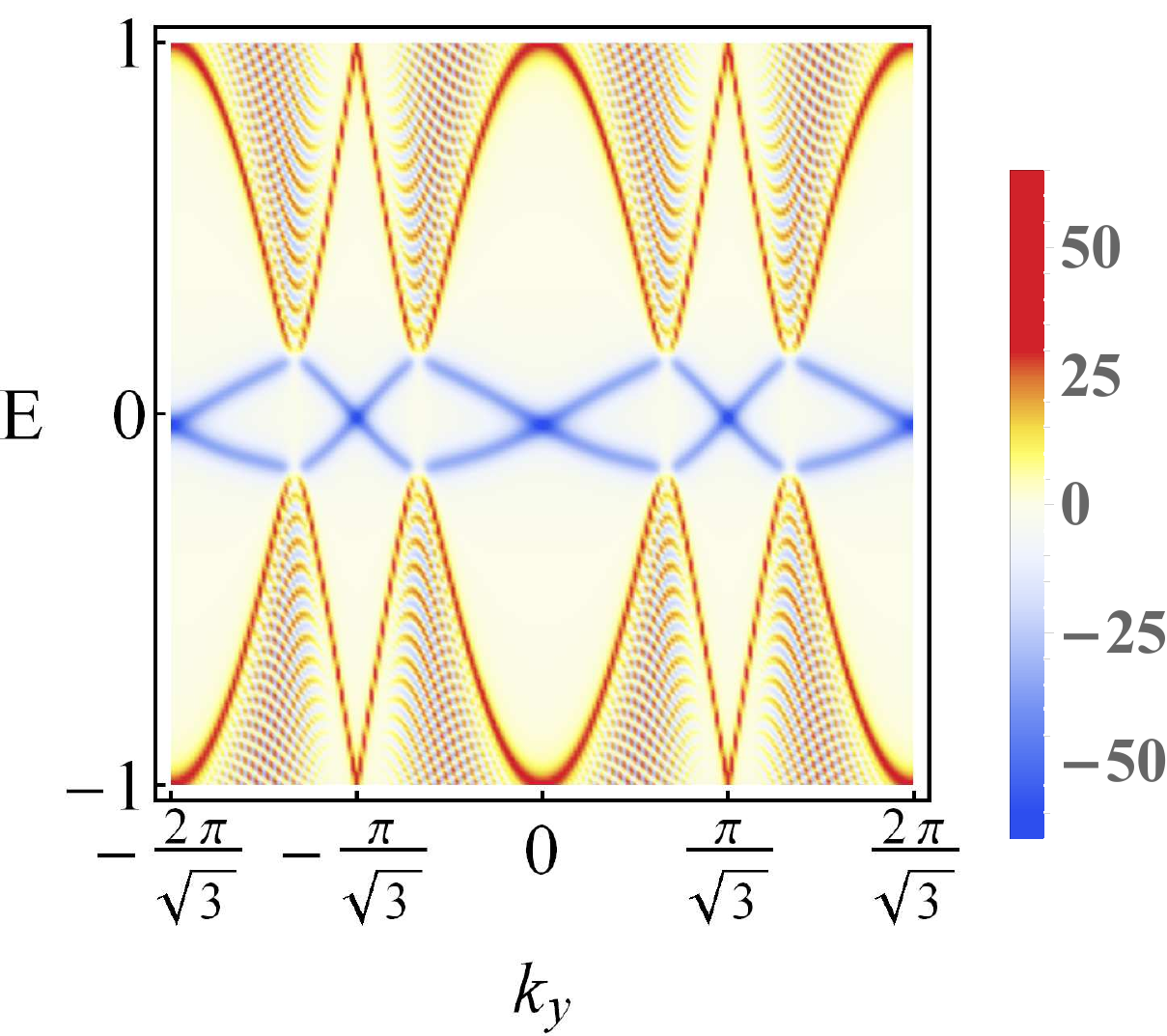}
    \end{minipage}
\caption{Left panel: energy spectrum obtained by an exact diagonalization of the Hamiltonian in Eq.~(\ref{eq:hamiltonian}) defined on a strip with one zigzag edge and one bearded edge. Right panel: the correction to the averaged spectral function due to the line impurity localized on either of the two sublattices. We consider $U=100$. For both plots, hopping amplitudes are taken to be $t=1, t_2=0.03$}
\label{fig:KMzigzagbearded}
\end{figure}

The formation of both zigzag and bearded edges can be recovered using our method by applying either of the potentials given in the left and middle panels of Fig.~\ref{fig:KMimpurities}. Fig.~\ref{fig:KMzigzagbearded} shows the energy spectrum obtained by exact diagonalization of the Hamiltonian on a strip with one zigzag edge and one bearded edge, along with the correction to the averaged spectral function due to a line impurity localized on either one of the two sublattices. Here we have enlarged the horizontal axis to $k_y \in [-2\pi/ \sqrt{3}; 2\pi/ \sqrt{3}]$ to see the edge states more clearly. We recover the same dispersion as in the main text for the zigzag edge states and obtain in addition other subgap states which are localized on the bearded edge. We need however to keep in mind the fact that in this case the two semi-infinite systems are not fully decoupled since the bulk Hamiltonian of the Kane-Mele model contains spin-flip NNN terms: this leakage effect needs to be taken into account carefully especially when we consider the spin-properties of the systems, but will not affect the spectrum below.

\end{document}